\documentclass[]{aastex631}
\usepackage{amsmath}
\usepackage[utf8]{inputenc}
\usepackage{graphicx}
\usepackage{enumerate}

\shorttitle{Relation of Observable Stellar Parameters to the Mass-Loss Rates of AGB Stars in the LMC}
\shortauthors{Prager et al.}
\graphicspath{{./}{figures/}}

\begin{document}
	
	\title{Relation of Observable Stellar Parameters to Mass-Loss Rate of AGB Stars in the LMC}
	
	\author[0000-0002-0658-6175]{Henry A. Prager}
	\affiliation{New Mexico Institute of Mining and Technology \\ Department of Physics
		\\ 801 Leroy Place \\
		Socorro, NM 87801, USA}
	\affiliation{Los Alamos National Laboratory
		\\ Los Alamos, NM 87545}
	\author[0000-0001-7921-8739]{Lee Anne Willson}
	\affiliation{Iowa State University \\ Department of Physics and Astronomy \\
		2323 Osborn Drive \\
		Ames, IA 50011, USA}
	\author[0000-0001-9910-9230]{Massimo Marengo}
	\affiliation{Iowa State University \\ Department of Physics and Astronomy \\
		2323 Osborn Drive \\
		Ames, IA 50011, USA}
	\author[0000-0002-8349-9366]{Michelle J. Creech-Eakman}
	\affiliation{New Mexico Institute of Mining and Technology \\ Department of Physics \\
		801 Leroy Place \\
		Socorro, NM 87801, USA}
	\begin{abstract}
		Using the \citet{2012ApJ...753...71R} data set for 6,889 pulsating AGB stars in the LMC, we have derived formulae for mass-loss rate as a function of luminosity and pulsation period or luminosity and mass in three ways, for each of five subsets of data: fundamental mode oxygen rich stars, first overtone mode oxygen rich stars stars, fundamental mode carbon stars, first overtone mode carbon stars, and extreme carbon stars.
		\textcolor{black}{Using the distribution of the stars in period versus luminosity and mass versus luminosity, we are able to derive a power-law fit to the dependence of mass-loss rate on those quantities.
			This results in formulae that reproduce observed mass-loss rates and are in general agreement with the expectation from mass-loss models that the mass-loss rate is highly sensitive to luminosity, mass, and pulsation period.}

		In the process of carrying out this analysis we have found radius-mass-luminosity and \textcolor{black}{examined} pulsation-mass-radius relations using published evolutionary and pulsation models. 
		These allow us to derive mass and radius from the observed quantities luminosity and pulsation period. 
		We also derived new mass-loss rate versus color relations.
	\end{abstract}
	

	\section{Introduction} \label{sec:intro}
	Asymptotic giant branch (AGB) stars have inert carbon/oxygen 
	cores with concentric hydrogen and helium-burning shells, surrounded by an expanded stellar envelope (see \citet{2005ARA&A..43..435H} and references therein).
	Near the tip of the AGB, hydrogen and helium burning alternate in thermal pulses, also called helium shell flashes; this produces modulation in the luminosity of the stars over 10,000 to 100,000 year cycles.
	Near the tip of the AGB the stars become unstable to pulsation, becoming Mira or semi-regular variables with large \textcolor{black}{variations in visual magnitude, radius, and luminosity with pulsation periods of hundreds of days.}
	The pulsation, combined with the formation of dust, drives a high mass-loss rate, removing most of the envelope in a mere 200,000 years \citep{2000ARA&A..38..573W}.
	When the envelope mass gets small enough, the star shrinks as it removes the residual envelope material, and what was the core becomes a new white dwarf star.
	Mass loss near the tip of the AGB is a major source of interstellar dust, and it is the mechanism by which stellar material, enriched with \emph{s}-process elements and carbon, returns to the interstellar medium
	(see reviews by \citet{2000ARA&A..38..573W} and  \citet{2018A&ARv..26....1H} for greater detail).
	
	\textcolor{black}{
		The first widely used mass loss formula for AGB stars was derived by \citet{1975psae.book..229R} from a small sample of red giants and supergiants, including a few AGB stars.
		As more data on M giant stars (on the red giant branch\textendash RGB\textendash and the asymptotic giant branch) became available, it was evident that this formula over predicted the mass-loss rates of RGB stars and under predicted the mass-loss rates in AGB stars, especially at the end of their lifetimes \citep{1978A&A....70..227K}.
		A number of other formulae were proposed based on observations or modeling of the mass loss process: \citet{1983AandA...127...73B,1988ApJ...331..435V,1990ApJ...365..301B,1995AandA...297..727B,2005AandA...438..273V,2002AandA...384..452W,2005ApJ...630L..73S}.
		A similar small-exponent dependence on parameters was found also by \citet{2009AJ....137.4810S} and \citet{2017MNRAS.465..403G}, but as \citet{2012ApJ...753...71R} note the scatter in this  data is so large that luminosity dependence cannot be estimated using standard methods.
		Alternative approaches have been taken by \citet{1993ApJ...413..641V} and \citet{2010AandA...523A..18D}, fitting the mass-loss rate as an exponential function of period alone, yielding significantly better fits to model data than the Reimers-like formulas.
		\citet{1993ApJ...413..641V}’s exponential formula produces a much steeper dependence on stellar parameters; steeper dependence is also supported by \citet{2010MNRAS.408..522K}, based on models of stellar clusters that needed steeper mass-loss dependence to match their observed evolutionary histories, and by \citet{2015A&A...581A..60D}, who found $\dot{M} \propto L_{\star}^{5}$.
	}
	
	Models for this mass-loss process are challenging because it is characterized by feedback between dynamics and grain growth in a low-density atmosphere that cannot be treated with equilibrium physics \citep{1988ApJ...329..299B,2000ARA&A..38..573W,2018A&ARv..26....1H}. 
	Several different mass-loss modeling codes are in use, each making different approximations or assumptions about the details.
	Overall, the results are similar: neither pulsation nor dust alone are capable of producing the high mass-loss rates that are observed, but they can do so when treated together \citep{2000ARA&A..38..573W,2018A&ARv..26....1H}.
	\textcolor{black}{In one dimension, the most detailed physical models are based on the DARWIN code \citep{2003A&A...399..589H, 2016A&A...594A.108H}.}
	\textcolor{black}{These models continue to become more sophisticated, such as through incorporating aluminum oxides and gradual iron enrichment \citep{2016A&A...594A.108H, 2022A&A...657A.109H}. 
		However, significant gaps in our} \textcolor{black}{understanding of mass loss} remain, including applicability in low metallicity stars \textcolor{black}{on the AGB \citep{2018MNRAS.481.4984M} to the far more massive red supergiant stars \citep{2022arXiv220502207B}.}
	\textcolor{black}{Most models also do not yet incorporate a full range of known physical effects such as dust-gas drift, departure from local thermodynamic and/or chemical equilibrium, or purely multi-dimensional processes such as convection \citep{2018A&ARv..26....1H}. }
	
	The primary goal of this work is to constrain the exponents \textcolor{black}{($B$, $C$) for} a power-law approximation \textcolor{black}{$\dot{M} = A L^B M^C$} for a \textcolor{black}{large,
		homogeneous} sample of stars in the LMC, using a subset of the \textcolor{black}{data set} by \citet{2012ApJ...753...71R}, with 
	the selection process described in Section \ref{sec:photometric_catalog}.
	\textcolor{black}{This addresses the discrepancy in luminosity dependence of mass-loss rates as measured by CO lines and as measured using dust, as noted in \citet{2018A&ARv..26....1H}.}
	\textcolor{black}{
		We also derive limits on the lead coefficient A, which determines at what luminosity this pattern appears. 
		$A$ is sensitive to some parameters that are less certain, but can be determined to be within a narrow range from the observations.
	}
	
	\textcolor{black}{To examine the evolutionary behavior and compare with previous mass loss formulae} we also needed to derive masses and radii. To do this we use evolutionary tracks to get the radius $R$ as a function of luminosity $L$ and mass $M$ (see \textcolor{black}{S}ection \ref{sec:RML_relations}), and pulsation studies to get the pulsation period $P$ as a function of $M$ and $R$ (see \textcolor{black}{S}ection \ref{sec:PMR_relations}).
	
	\section{Methodology} \label{sec:methodology}
	
	\textcolor{black}{
		We define the critical mass-loss rate $\dot{M} = M/t_{\mathrm{ev}}$
		\begin{equation}
			\frac{1}{t_{\mathrm{ev}}} \equiv \frac{1}{L} \frac{\mathrm{d}L}{\mathrm{d} t} = \frac{\mathrm{d} \ln L}{\mathrm{d} t}.
		\end{equation}
		To derive $t_{\mathrm{ev}}$, we look at the core mass\textendash luminosity relations.
		These relations are discussed by \citet{1988ApJ...328..641B}, who summarize prior work, and more recently by \citet{2019MNRAS.482..929T}.
		For a relation of the form $L = C_{1}(M_{c}-C_{2})$\textcolor{black}{, where $L$ is the luminosity and $M_{c}$ is the core mass,} we have $\mathrm{d} L/\mathrm{d} t = C_{1} \mathrm{d} M_{c}/\mathrm{d} t$ and \textcolor{black}{an energy conversion rate}  $L = \textcolor{black}{0.006} c^2 \mathrm{d} M_{c}/\mathrm{d} t$ \textcolor{black}{(where $c$ is the speed of light) \citep{2013sse..book.....K}}, giving
		\begin{equation}
			t_{\mathrm{ev}} = \textcolor{black}{\frac{8.84 \times 10^{10}}{C_{1}}\ \mathrm{yr}}
		\end{equation}
		with $C_{1}$ in solar units.
		The relations summarized in the above cited papers tell us $t_{\mathrm{ev}}$ is between \textcolor{black}{$1.3$ and $1.6\ \mathrm{Myr}$}, with a hint that lower metallicity stars will have slower evolution.
		We conclude that $t_{\mathrm{ev}}$ is likely between \textcolor{black}{1.2 and 1.7 Myr} for the LMC stars. 
		We have chosen to set $\textcolor{black}{\log(}1/t_{\mathrm{ev}}\textcolor{black}{)} = \textcolor{black}{\log (}\dot{M}_{\mathrm{crit.}}/M\textcolor{black}{)}=-6.2$, equivalent to $t_{\mathrm{ev}}=10^{6.2}\ \mathrm{yr} \approx \textcolor{black}{1.58}\ \mathrm{Myr}$. 
		\textcolor{black}{Note that the evolution time is not the same as the AGB lifetime.
			It is a characteristic time that describes the growth of luminosity over the lifetime of the star.
		}
		From $t_{\mathrm{ev}}$, we can define the critical mass-loss rate, or the death line, where luminosity-dominated evolution shifts to mass-dominated evolution.
		\begin{equation}
			\left(\frac{1}{M} \frac{\mathrm{d} M}{\mathrm{d} t}\right)_{\mathrm{crit.}} = \frac{\dot{M}_{\mathrm{crit.}}}{M} = \frac{1}{L} \frac{\mathrm{d} L}{\mathrm{d} t} = \frac{1}{t_{\mathrm{ev.}}}
		\end{equation}
		In earlier work (\emph{e.g.}, \citealt{1991ApJ...375L..53B}), the death line was sometimes referred to as ``the (AGB) cliff''. 
		This will be further discussed and used in Section \ref{sec:mdot_as_f_L_M}.
	}
	
	We have examined two independent methods for constraining the exponents: (a) bilinear fits to the mass-loss rate, luminosity, and pulsation period (see \textcolor{black}{S}ection \ref{sec:mdot_as_f_L_P}) and (b) fitting the slope and width of the distribution and mass-losses of stars in luminosity\textendash pulsation period space (see \textcolor{black}{S}ection \ref{sec:mdot_as_f_L_M}).
	We will see that direct linear fits using $\dot{M}$ result in low reliability as we move further from the mean mass-loss rate, fits with $L$ as the dependent variable are unreliable due to high scatter in that parameter, and that formulas found using the \textcolor{black}{PL} strip result in the overall best fits with similar reliability throughout the range of mass-loss rates while having tolerable amounts of spread.
	Finally, we will examine these results in the context of prior mass-loss formulae and the limitations imposed by observation and the methods used (see \textcolor{black}{S}ection \ref{sec:comparison_with_other_formulae}).
	In the process of this work, we have also determined new mass-loss rate\textendash color relations, and these can be found in Appendix \ref{sec:mdot_from_color}. 
	
	
	\textcolor{black}{
		The primary limitation of this study is that it has only been done for one metallicity, assumed appropriate for all of the LMC stars. 
		We have also only considered two bins of $C/O$, while the mass-loss rates may be sensitive to the values \textcolor{black}{within} each bin. }
	
	\textcolor{black}{
		We have used a fixed dust-to-gas ratio to translate observed dust mass-loss rates to total mass-loss rates, based on estimates from the literature. 
		If our value is incorrect but there is a single value that works for all mass-loss rates, the effect will be to shift the pattern without changing the derived exponents. 
		However, the dust-to-gas ratio is potentially sensitive both to $\mathrm{C}/\mathrm{O}$ and to the mass-loss rate; this could affect the exponents in the mass-loss formula.
		There is also the potential for a mass-loss dependent bias due to fixed expansion velocities \citep{2018MNRAS.481.4984M}.
	}

	\section{Photometric Catalog} \label{sec:photometric_catalog}
	For this analysis, we are using the data compiled in \citet{2010ApJ...723.1195R} and \citet{2012ApJ...753...71R}; the authors have fitted models from the ``Grid of Red Supergiant and Asymptotic Giant Branch ModelS (GRAMS)'' \citep{2011ApJ...728...93S,2011A&A...532A..54S} to all known AGB and red supergiant (RSG) stars in the Large Magellanic Cloud, combining results from the photometric Magellanic Clouds Photometric Survey (MCPS) \citep{2004AJ....128.1606Z}, Two Micron All Sky Survey (2MASS) survey \citep{2006AJ....131.1163S}, and Surveying the Agents of a Galaxy's Evolution (SAGE) \citep{2006AJ....132.2268M} survey with the variability data from the MAssive Compact Halo Objects (MACHO) survey \citep{1997ApJ...486..697A} \textcolor{black}{which have been refitted by \citet{2008AJ....136.1242F}}.
	Because the \citet{2012ApJ...753...71R} data set includes RSG, AGB, and other long-period variable stars, we sorted out the AGB stars from the rest of the data set.
	We can distinguish AGB stars from the others based on their brightness, pulsation periods, and colors.
	
	\textcolor{black}{We are particularly interested in the behavior of stars that are near the death line (defined in \textcolor{black}{S}ection \ref{sec:intro}).
		The evolution time $1/t_{\mathrm{ev}}=L^{-1} \mathrm{d} L/\mathrm{d} t\ \textcolor{black}{= \mathrm{d} \ln L/\mathrm{d} t}$ is constant for a star on the AGB \textcolor{black}{with a core mass-luminosity relation as posited in Section \ref{sec:methodology}}, \textcolor{black}{such} that $\Delta \log L\ \textcolor{black}{\approxeq}\ 2.3 \Delta t/t_{\mathrm{ev}}$.
		The death line or cliff is defined by where $\dot{M} = \dot{M}_{\mathrm{crit.}} = M/t_{\mathrm{ev}}$, and can be \textcolor{black}{demonstrated} in \textcolor{black}{either} $M(t)$ or $\log M$ vs. $\log L$.
		The region around the death line ($0.1 \dot{M}_{\mathrm{crit.}} < \dot{M} < 10 \dot{M}_{\mathrm{crit.}}$) is the ``death zone''.
		The death zone is where AGB stars shift from luminosity-dominated evolution to mass-dominated evolution.
		Thus, this analysis does not attempt to include stars with the very highest mass-loss rates ($\dot{M} \gtrsim 2 \times 10^{-5}\ \mathrm{M_{\sun}/yr})$ or the longest periods (the longest period in the selected set is $817.7\ \mathrm{days}$), where the models show that the physics of the mass loss process may fundamentally change \citep{2000ARA&A..38..573W,2018A&ARv..26....1H}.
	}
	
	\textcolor{black}{The GRAMS dust mass-loss rates have some limitations.
		As described in detail in \citet{2012ApJ...753...71R}, the grid assumes spherical symmetry, fixed expansion velocities, dust compositions, and optical constants. 
		These assumptions may lead to systematic errors in the mass-loss rates.
		Of particular note, the effects of optical constants are particularly strong in the carbon stars \citep{2018A&A...609A.114G} and the fixed expansion velocities potentially cause a mass-loss dependent bias \citep{2018MNRAS.481.4984M}.
		However, this data set is exceptional in its size, being an order of magnitude larger than what we have for local AGB stars, making it ideal for the population-scale analysis seen in Section \ref{sec:mdot_as_f_L_M}.
	}
	
\begin{figure*}[ht]
		\centering
		\includegraphics{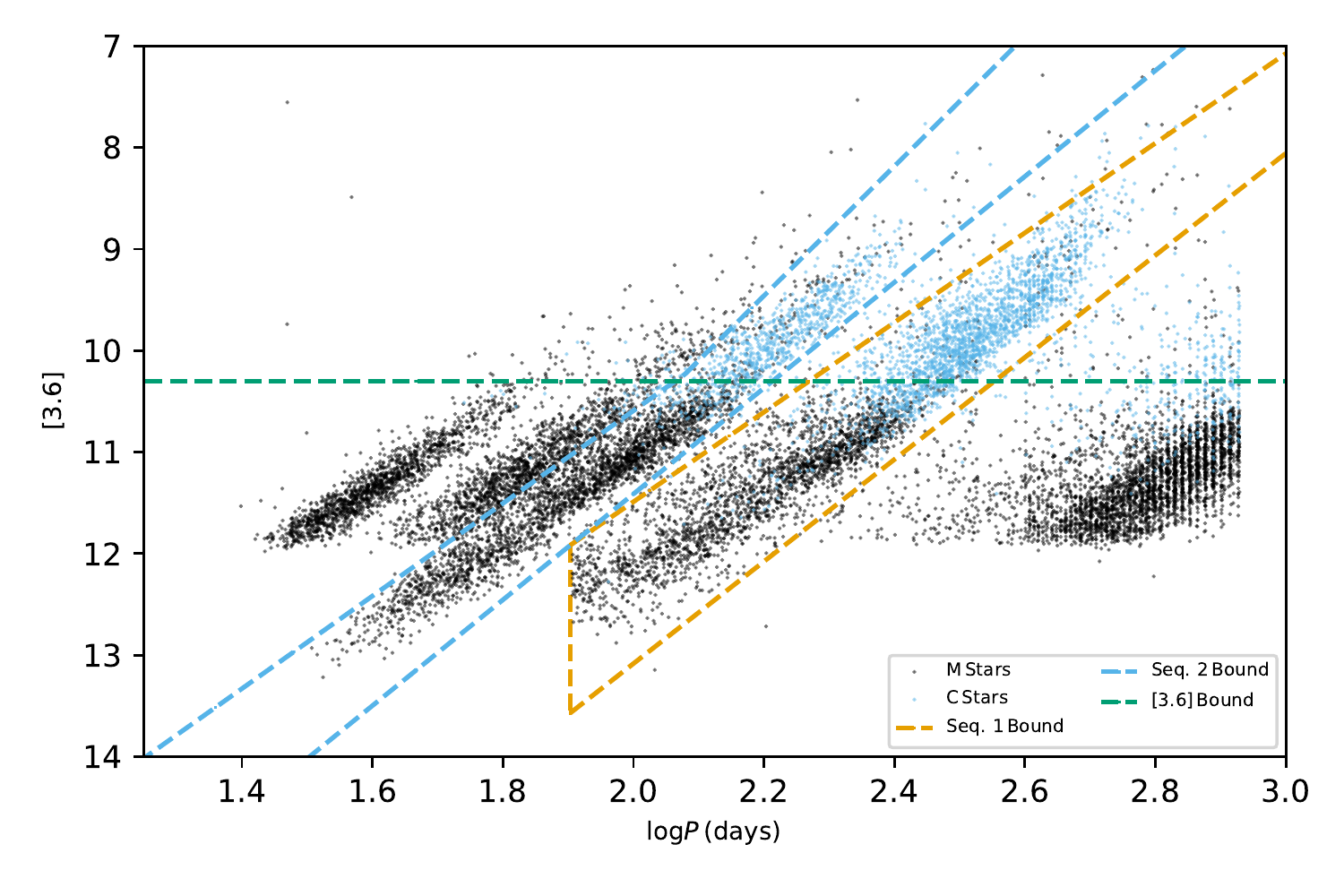}
		\caption{Plot of [3.6] vs $\log P$ for the stars in the \citet{2012ApJ...753...71R} data set. M-type stars are in black and C-type stars in blue. 
			M stars brighter than $[3.6]=10.3$ are assumed to be red supergiant stars, and are excluded from the analysis. 
			Following the naming convention established in \citet{2010ApJ...723.1195R}, sequence 1 defines AGB stars pulsating in the fundamental-mode, and sequence 2 defines AGB stars pulsating in the first overtone-mode. Definitions of the sequence bounds can be found in equations \ref{eq:seq1_right}\textendash \ref{eq:seq2_left}.}
		\label{fig:sorting_of_AGB_stars}
	\end{figure*}

	We are most interested in AGB stars pulsating in the fundamental and first overtone modes.
	In a period-magnitude diagram, these stars are found in sequences 1 and 2, as defined by \citet{2010ApJ...723.1195R} and following previous works: \citet{1999IAUS..191..151W,2004MNRAS.347..720I,2005AJ....129..768F}; and \citet{2009MNRAS.395L..11G}.
	To efficiently pick stars in these sequences out, we defined them in log-log space using $3.6\ \mathrm{\mu m}$ flux, converted to magnitude using the photometric zero point of $F=280.9\ \mathrm{Jy}$ \citep{2004ApJS..154...10F}, and pulsation period ($P$) to define linear bounds.
	\begin{align}
		\text{Seq. 1, Right: } [3.6] &= 6.97(\log P - 2.51) + 9.25 \label{eq:seq1_right} \\
		\text{Seq. 1, Left: } [3.6] &= 3.68(\log P - 2.73) + 9.44 \label{eq:seq1_left}\\
		\text{Seq. 2, Right: } [3.6] &= 5.21(\log P - 2.57) + 8.42 \label{eq:seq2_right} \\
		\text{Seq. 2, Left: } [3.6] &= \begin{cases}
			4.56(\log P - 2.12) + 10.04 & \text{ if } P \leq 120 \\ 
			6.39(\log P -2.30) + 8.81 & \text{ if } P > 120 
		\end{cases} \label{eq:seq2_left} 
	\end{align}
	These bounds and the stars found within are depicted in Figure \ref{fig:sorting_of_AGB_stars}.
	RSGs are also present in this data set, so to limit their presence as much as possible, we cut all M stars brighter than $[3.6]=10.3\text{ mag}$ and $\log L>4.5$.
	\textcolor{black}{The stars with relative errors $>1$, or no reported error in $\dot{M}$ or $L$, have been removed as well.
		We have also removed stars with} \textcolor{black}{a derived mass} $M<0.45\ \mathrm{M_{\sun}}$\textemdash below the lowest final mass predicted by initial-final mass relations \citep{2014ApJ...782...17K, 2012ApJ...746..144Z}. \textcolor{black}{These masses were derived using the relations established in Section \ref{sec:rm_model_grids}.}
	
\begin{figure*}[ht]
		\gridline{
			\fig{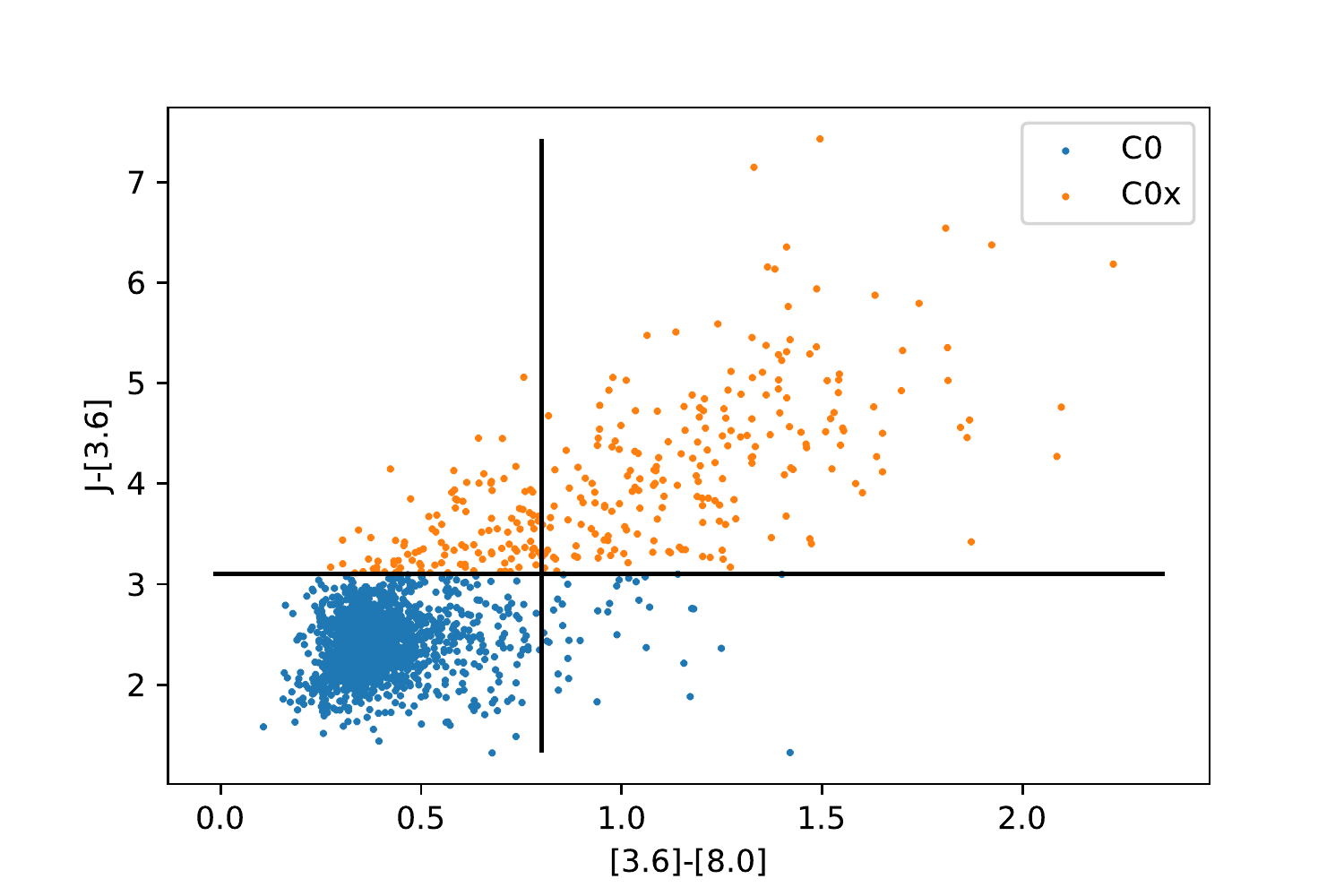}{0.49\textwidth}{(a)}
			\fig{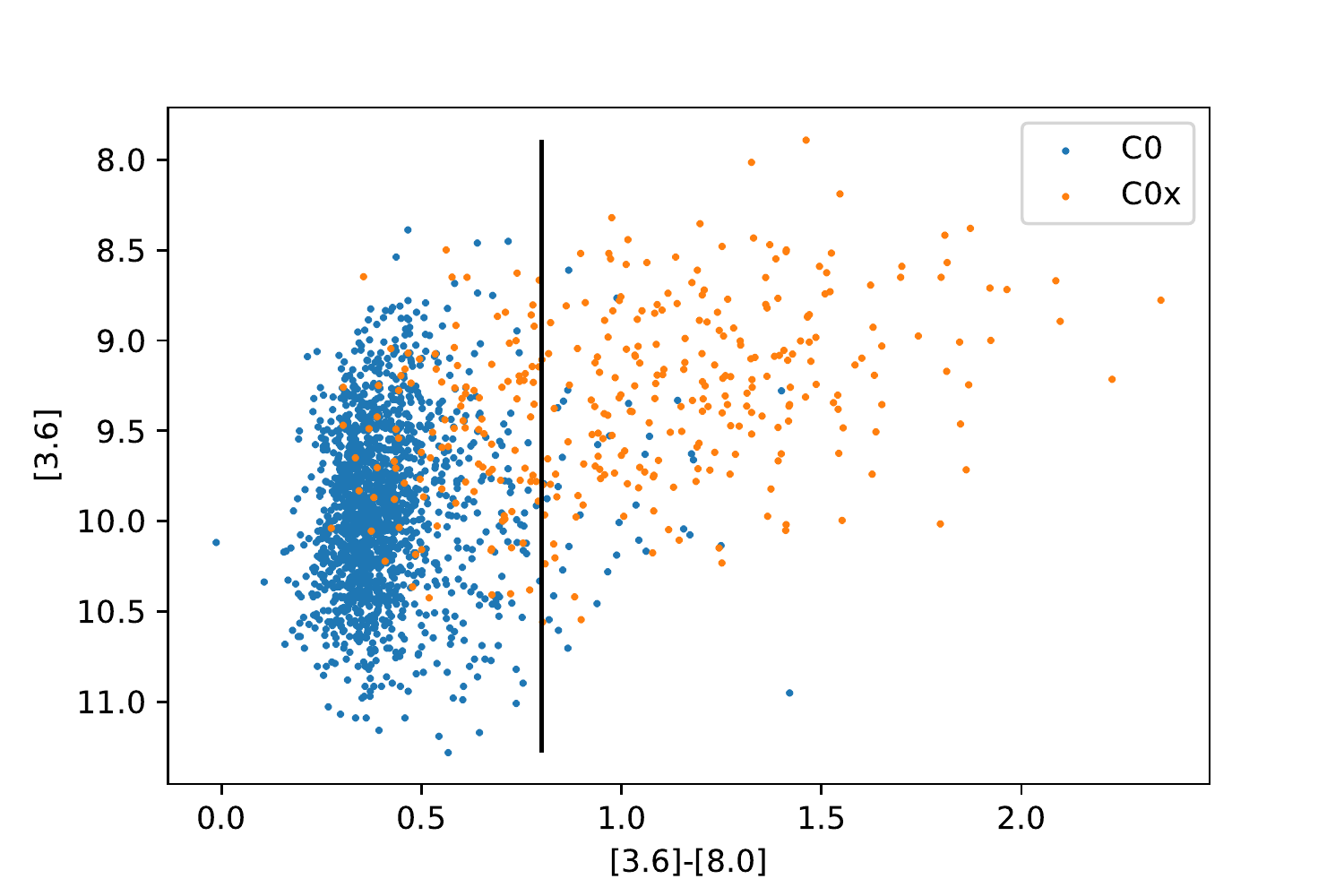}{0.49\textwidth}{(b)}
		}
		\caption{The split of carbon-rich, fundamental-mode pulsator (C0) AGB stars into `normal' and `extreme' populations is depicted here, using the method found in \citet{2006AJ....132.2034B, 2009AJ....137.4810S}; and \citet{2011AJ....142..103B}, where xAGB stars are those brighter than the $3.6\ \mathrm{\mu m}$ TRGB and with $J -[3.6] > 3.1$ mag (a), or if there is no near-IR detection, $[3.6]-[8.0] > 0.8$ (b). Stars in blue are `normal' C-stars, and stars in orange are `extreme' AGB stars. All stars were able to be sorted through the primary method, so the scattered blue stars in figure (b) may be ignored.}
		\label{fig:C0_split}
	\end{figure*}

	AGB stars can be divided into two types based on the ratio of carbon to oxygen: $C/O <1$ (M stars) and $C/O>1$ (carbon \textendash\ C \textendash\ stars).
	In the LMC, the stars can be distinguished by color and magnitude (see \citet{2006AJ....132.2034B} for details), and have already been separated as such in the \citet{2012ApJ...753...71R} data set.
	We have continued to use this classification here. 
	Further, the carbon stars pulsating in the fundamental-mode undergo an extreme mass-loss phase, so they can be split into ``normal'' AGB stars and ``extreme'' (xAGB) stars.
	To separate the two populations, we followed the procedures found in \citet{2006AJ....132.2034B,2009AJ....137.4810S}; and \citet{2011AJ....142..103B}, among others.
	In these papers, xAGB stars are defined as fundamental mode carbon stars brighter than the $3.6\ \mathrm{\mu m}$ tip of the red giant branch (TRGB) and with $J -[3.6] > 3.1$ mag, with a fallback classification of $[3.6]-[8] > 0.8$ for stars brighter than the $3.6\ \mathrm{\mu m}$ TRGB with no near-infrared observations.
	In the \citet{2012ApJ...753...71R} data set, nearly all carbon stars have measurements in the $J$ band and in the $3.6\ \mathrm{\mu m}$ band, and are thus classified using those measurements (Figure \ref{fig:C0_split}).
	The final results of the classification of stars based on sequence and \textcolor{black}{$C/O$} composition can be found in Table \ref{tab:stellar_stats}, \textcolor{black}{along with the mean standard deviation of the mass-loss rates in the final column}.

	\begin{deluxetable}{ccccccc}[h]
		
		
		
		
		\tablecaption{Statistics of Analyzed AGB Stars}
		
		
		\tablehead{\colhead{C/O Ratio} & \colhead{Pulsation Mode} & \colhead{Notes} & \colhead{Identifier} & \colhead{Count} & \colhead{\textcolor{black}{Min., Median, Max. ($\log \dot{M}$)}} & \colhead{\textcolor{black}{$\sigma_{\log \dot{M}}$}}} 
		
		\startdata
		$<1$ &  Fundamental &   &  M0 &  \textcolor{black}{1979} & \textcolor{black}{-7.83, -6.42, -5.14}& \textcolor{black}{0.39}\\
		$<1$ &  First Overtone &   &  M1 &  2162 & \textcolor{black}{-7.97, -6.57, -4.96 }& \textcolor{black}{0.44}\\
		$>1$ &  Fundamental &   &  C0 &  1654 & \textcolor{black}{-7.67, -6.62, -5.44}& \textcolor{black}{0.25}\\
		$>1$ &  Fundamental &  Extreme AGB &  C0x &  341 & \textcolor{black}{-6.49, -5.74, -4.67}& \textcolor{black}{0.34}\\
		$>1$ &  First Overtone &   &  C1 &  781  & \textcolor{black}{-7.68, -6.56, -4.33}& \textcolor{black}{0.29}\\
		\enddata
		
		
		\tablecomments{\textcolor{black}{Properties and identifiers for our five AGB star categories.
				As we will see in Section \ref{sec:mdot_from_parameters}, the C0x population is a continuation of the C0 population.}}
		
		
	\end{deluxetable}\label{tab:stellar_stats}
	\newpage
	The pulsation period $P$ is well established, but the luminosity $L$ is subject to significant scatter.
	This scatter limits our ability to narrow the dependence of $\dot{M}$ on $L$ using these observations \citep{2012ApJ...753...71R, 2018A&ARv..26....1H}, and we will see this appearing as regression dilution  \textcolor{black}{\citep{10.2307/1412159,1999_10.2307/2680496,carroll2006measurement}} in the fits.
	To get around this issue, we will explore two complementary methods for constraining the mass-loss formulae using this data, in \textcolor{black}{S}ections \ref{sec:mdot_as_f_L_P} and \ref{sec:mdot_as_f_L_M}.
	When the color is taken into account, there is an observational distinction between the fundamental mode C stars and the extreme (redder) fundamental mode C0x stars, and so they are treated separately in Section \ref{sec:mdot_as_f_L_P} and Appendix \ref{sec:mdot_from_color}. 
	However, in \textcolor{black}{S}ection \ref{sec:mdot_as_f_L_M} it is clear that the physical properties\textemdash $L$, $M$, $P$ and $R$\textemdash do not separate these stars from the others, and so they are more appropriately binned together \textcolor{black}{and understood to be two parts of the same population}.
	
	\subsection{Dust-to-Gas Ratio and Mass-Loss Rates} \label{sec:dust-to-gas}
	Dust mass-loss rates are provided in the \citet{2012ApJ...753...71R} database, based on fitting to models in the GRAMS grid.
	For ease of comparison to other works, we want the total mass-loss rate of the stars.
	However, the exact relation between the dust mass-loss rate, $\dot{M}_{\mathrm{dust}}$, and the gas mass-loss rate, $\dot{M}_{\mathrm{gas}}$, (and combined, the total mass-loss rate $\dot{M}$) depend on factors that are not well known \citep{2000A&A...354..125V}.
	\textcolor{black}{Therefore, we will select values based on the final equations in \citet{2000A&A...354..125V}, but will also examine the consequences if the actual ratio differs from these values in Section \ref{sec:results}}
	\begin{align}
		\log \psi_{M} = &0.7^{+0.6}_{-0.3} \log Z/Z_{\sun} + \mathrm{constant}_{M} \label{eq:van_loon_dust_to_gas_M}\\
		\log \psi_{C} = &1.0^{+0.3}_{-0.3} \log Z/Z_{\sun} + \mathrm{constant}_{C}. \label{eq:van_loon_dust_to_gas_C}
	\end{align}
	Here, $\psi$ is the dust-to-gas ratio and $Z$ is the metallicity of the star.
	The constants in these equations are unknown, but we know approximately the value for stars of solar metallicity: 500 and 200, respectively \citep{1978ppim.book.....S,2012A&A...537A.105G,2012ApJ...753...71R}, which are noted here as well as in \citet{2008A&A...487.1055V} as being a poor fit for the low metallicity stars in the LMC.
	Using these values to calibrate, we found $1/\psi_{M} \approx  2913$ \textcolor{black}{($\mathrm{constant}_{M} = -\log 500$)} and $1/\psi_{C} \approx 2480$ \textcolor{black}{($\mathrm{constant}_{C} = -\log 200$)} \textcolor{black}{while using $Z=0.001$, consistent with the star formation history found by \citet{2009AJ....138.1243H} and what seems to be necessary to replicate the total mass-loss rates derived by \citet{2020MNRAS.498.3283P} using these dust-to-gas ratios}.
	\textcolor{black}{These larger ratios also bring the mass-loss rates into the same range as seen in solar metallicity stars, consistent with the results of \citet{2018MNRAS.481.4984M} which see mass-loss as being essentially independent of metallicity.}
	This scaling can easily be undone or adjusted when this relationship is better understood. 
	
\begin{figure*}[h]
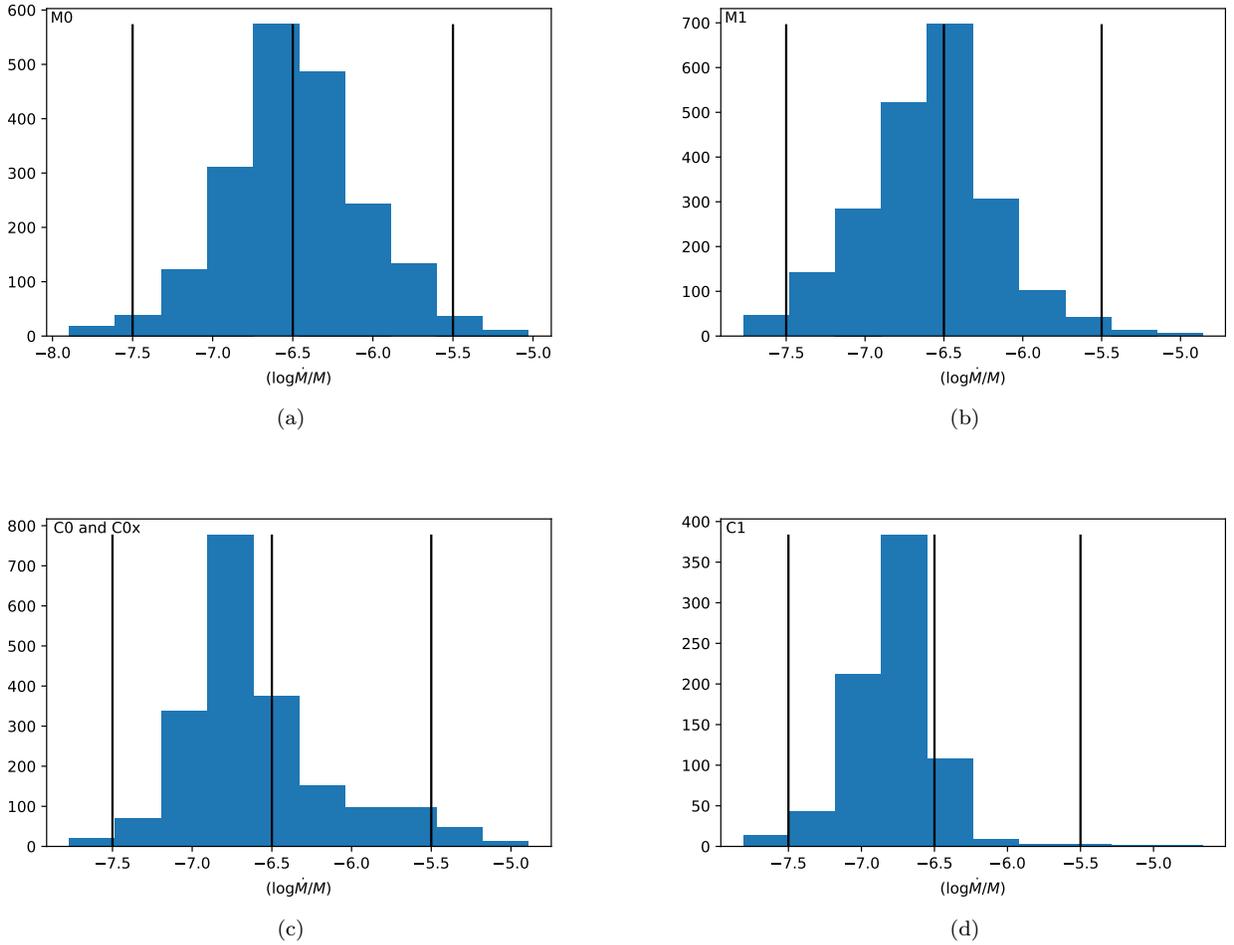

		\centering
		\gridline{
			\fig{figures/Death_Zone/mdot_per_m_histograms/logMdot_per_m_histogram_M0_data_tev_3162277.pdf}{0.48\textwidth}{(a)}
			\fig{figures/Death_Zone/mdot_per_m_histograms/logMdot_per_m_histogram_M1_data_tev_3162277.pdf}{0.48\textwidth}{(b)}
		}
		\gridline{
			\fig{figures/Death_Zone/mdot_per_m_histograms/logMdot_per_m_histogram_C0_and_C0x_data_tev_3162277.pdf}{0.48\textwidth}{(c)}
			\fig{figures/Death_Zone/mdot_per_m_histograms/logMdot_per_m_histogram_C1_data_tev_3162277.pdf}{0.48\textwidth}{(d)}
		}
		\caption{\textcolor{black}{The distribution of \textcolor{black}{$\log(\dot{M}/M)$} showing the death line ($\dot{M}/M = 1/t_{\mathrm{ev}}$) and the death zone ($\dot{M}/M = 1/t_\mathrm{ev} \pm 1$). The drop off above the death line results from rapid evolution at high mass-loss rates. 
				The drop off below the death line is presumed to be due to the difficulty of measuring lower dust mass-loss rates and/or a lower dust-to-gas ratio at low mass-loss rates.}
		}
		\label{fig:logMdot_per_m_histograms}
	\end{figure*}

	\textcolor{black}{
		Figure \ref{fig:logMdot_per_m_histograms} displays histograms of $\dot{M}/M$, chosen because $\dot{M}_{\mathrm{crit.}}=M/t_{\mathrm{ev}}$. 
		This lets us indicate the location of the death line in the samples.
		As explained in Section \ref{sec:intro}, we will be using an evolution time of $3.2\ \mathrm{Myr}$; thus, at the death line, $\log (\dot{M}/M)=-6.5$. 
		In all four cases, with this $t_{\mathrm{ev}}$ value, the stars are clustered close to the death line and a majority are found in the death zone (defined as $\pm 1$ dex in $\log \dot{M}/M$ from the death line).
		The decrease in the histogram for mass-loss rates above the critical mass-loss rate is expected from the more rapid evolution once $M$ is decreasing quickly.
		The decline on the low mass loss side is presumed to be the result of a combination of lower mass-loss rates producing less dust, lower mass-loss rates being harder to detect, and (given the intrinsic scatter in the mass-loss rates) our exclusion of stars with derived mass-loss rates below $10^{-8}\ \mathrm{M_{\sun}/yr}$.
		The lower mass-loss rate stars may also show lower amplitude light curves and miss being included because we only included stars with a measured period.
	}
	
	\textcolor{black}{
		The strong conclusion from these histograms is that in all four categories, the stars populate the death zone, and thus, a star does not need to be in a particular mode of pulsation to lose its envelope as an AGB star, though an overtone pulsator will do so at a slightly higher $L$.
		In fact, to reach and populate the overtone strip, a star must avoid being in the fundamental mode when it passes through the fundamental mode death zone, or it must avoid passing through the fundamental mode death zone, for example through an episode of rapid mass loss.
	}
	\newpage
	\section{Obtaining R and M Through Model Grids} \label{sec:rm_model_grids}
	
	For our bilinear fits, we will be using the observed quantities $\dot{M}$, $L$, and $P$. 
	\textcolor{black}{To put these results into the context of stellar evolution, we will need to derive the mass $M$ and radius $R$ for each star as well.} 
	Here, we find evolutionary tracks $R(L,M)$ and period-mass-radius relations $P(M,R)$.
	\textcolor{black}{We have $P(M,R)$ and  $R(L,M)$ relations (\emph{cf.} \citet{1984ApJ...277..333I,2019MNRAS.482..929T,1982ApJ...259..198F, 1986ApJ...311..864O}), but they either need to be re-assessed using modern results or utilize parameters that are simply not found in the \citep{2012ApJ...753...71R} data set.}
	We will need these variables and relations to analyze the distribution in $L$ and $M$ and for comparison to other formulae.
	\textcolor{black}{Going forward, we will be assuming that the photospheric radius (found from
		evolutionary models) and the pulsation radius (used in the period-mass-radius relation) coincide.}
	\subsection{Radius-Mass-Luminosity Relations} \label{sec:RML_relations}
	
	Evolutionary models tell us what radius and luminosity to expect for a star of mass $M$ and the fractional metallicity $Z$ at a given evolutionary stage.
	For AGB stars, the radius depends on the choice of mixing length parameter; for a grid of models this is usually tuned by forcing the models to fit the present day Sun and/or to produce isochrones that fit observations of clusters.
	Earlier work (e.g. \citet{1991ApJ...375L..53B}) used relationships from \citet{1984ApJ...277..333I}.
	For this work, we will use the PARSEC-COLIBRI isochrones \citep{2012MNRAS.427..127B, 2013MNRAS.434..488M, 2014MNRAS.445.4287T, 2020MNRAS.498.3283P} to find a set of evolutionary tracks $R(L,M)$ for a composition appropriate \textcolor{black}{for AGB stars in} the LMC, while also separating M-type and C-type stars.
	\textcolor{black}{We used isochrones in intervals of $\Delta \log t = 0.05$ between $\log t = 8.0$ and $\log t = 10.1$ years. 
		AGB stars were then selected by choosing isochrone stars with $\log (L/L_{\sun})$ between 2.9 and 4.2 that have a listed pulsation period or periods.}
	
\begin{figure*}[ht]
		\centering
		\gridline{
			\fig{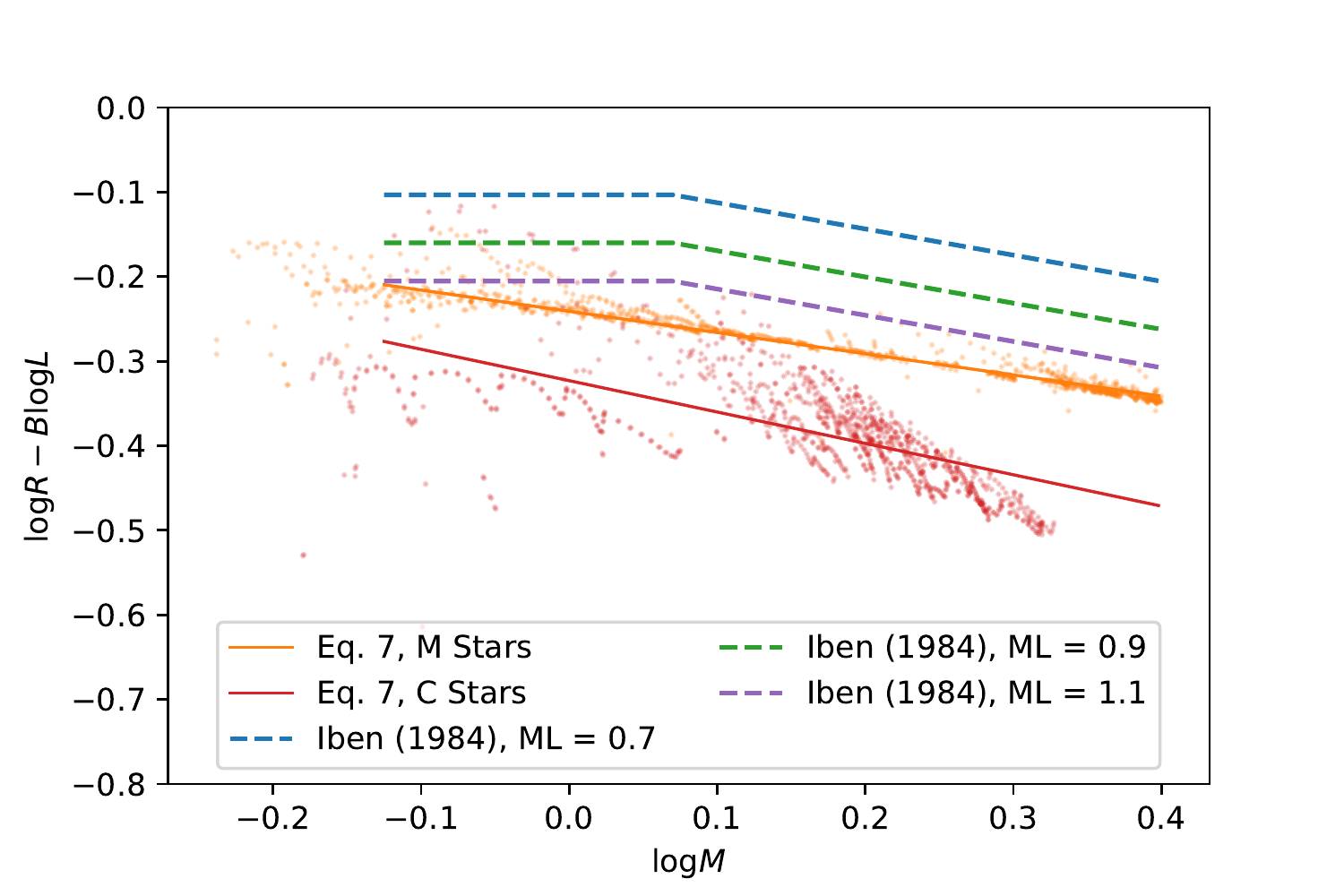}{0.49\textwidth}{(a)} 
			\fig{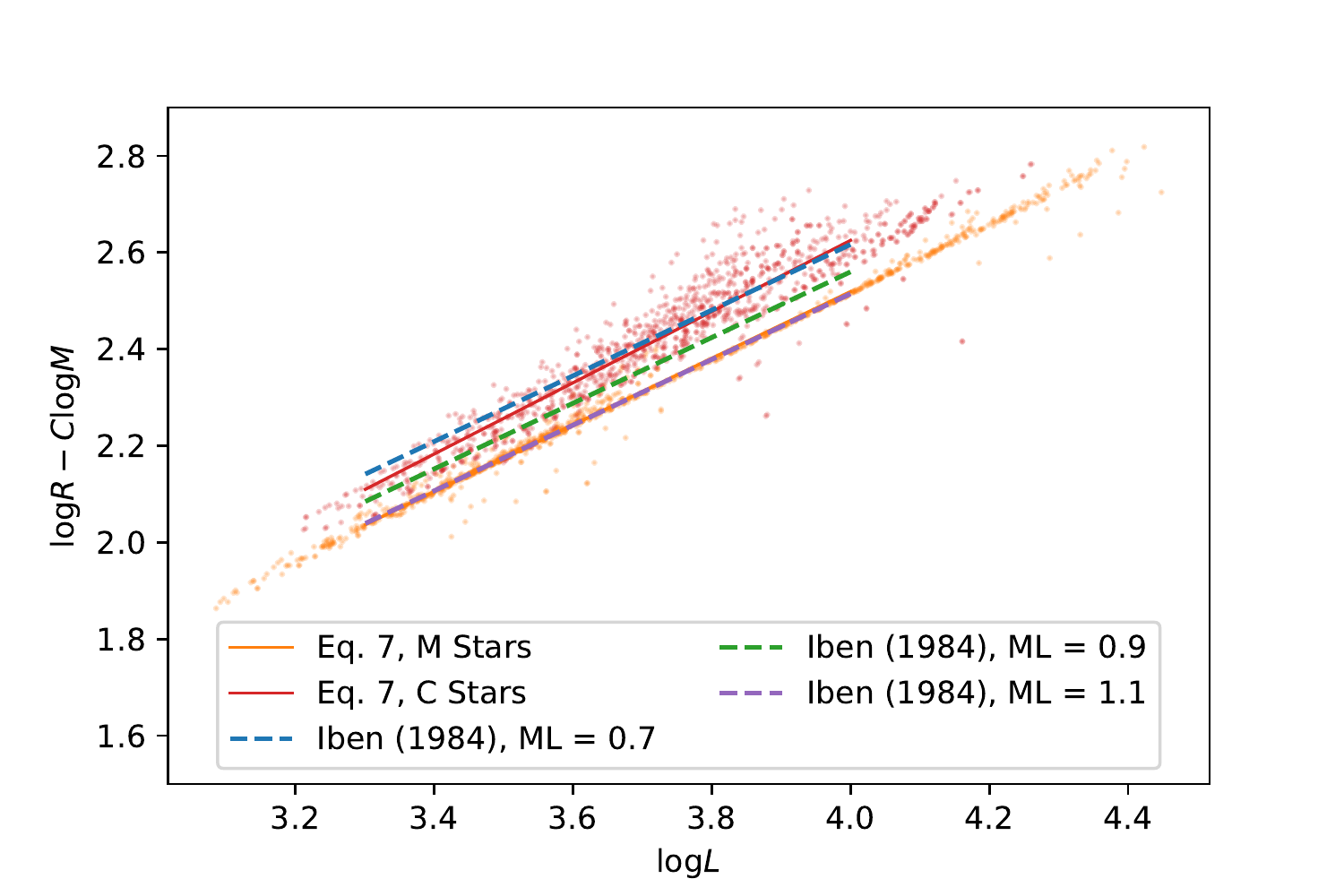}{0.49\textwidth}{(b)} 
		}
		\caption{\textcolor{black}{To project this three-dimensional fit into two dimensions,} (a) has had the luminosity dependence of the relations removed and (b) has had the mass dependence removed. Same color dots show the models that were used to determine the relations, color matched to the fits. }
		\label{fig:RML_comparisons}
	\end{figure*}

	For our analysis, we chose isochrones for stars with metallicity $Z=0.003$, similar to the metallicity found in the LMC 10 Gyr before present \textcolor{black}{(see \citealt{2009AJ....138.1243H}, Figure 11)}, representing a typical age for the stars in our sample.
	\textcolor{black}{We have not determined if there is a metallicity dependence in the radius-luminosity-mass relations, and leave that for a more dedicated analysis than what is here.}
	Using these isochrones, we found a power law fit to the equation:
	\begin{equation}
		\log R = \log A_{RLM} + B_{RLM} \log L + C_{RLM} \log M.\label{eq:evolutionary_track}
	\end{equation} 
	The results of these fits can be found in Table \ref{tab:RLM_fits}.
	\textcolor{black}{Prior work (e.g. \citet{1991ApJ...375L..53B}) used a formula from \citet{1984ApJ...277..333I} that had mixing length as a free parameter and a change of slope $\mathrm{d} \log R/\mathrm{d} \log M$ at $M=1.175$.
		Our new fits to the isochrone data are based on models with calibrated mixing length and \textcolor{black}{the above metallicity}, and show no discontinuity in the slope.}
	\textcolor{black}{Note that this metallicity is slightly different from the value used in Section \ref{sec:dust-to-gas} because this was the nearest value available in the TP-AGB isochrone grids when we did this work.}
	Graphical comparisons of our results to \citet{1984ApJ...277..333I} can be found in Figure \ref{fig:RML_comparisons}.
	Our relation for M stars agrees well with with the luminosity dependence of the \citet{1984ApJ...277..333I} relation when the mixing length is 1.1; the same is true for the C stars when mixing length is 0.7 or 0.9, depending on the luminosity. 
	The mass dependence is significantly different \textcolor{black}{between $C/O$} compositions.
	\begin{deluxetable*}{cccc}[h]
		
		
		
		
		\tablecaption{Fit of $\log R =  \log A_{R L M} + B_{R L M} \log L + C_{R L M} \log M$}
		
		
		\tablehead{\colhead{Composition} & \colhead{$\log A_{R L M}$} & \colhead{$B_{R L M}$} & \colhead{$C_{R L M}$}  } 
		
		\startdata
		M &   -0.241(9) & 0.690(3) & -0.251(4) \\
		C & -0.323(27) & 0.737(7) & -0.371(14) \\
		\enddata
		
		\tablecomments{$R$, $L$, and $M$ are measured in solar units. Values in parentheses are the uncertainty, using standard error. $Z$ was taken to be $0.003$, consistent with AGB stars \textcolor{black}{formed} in the LMC \textcolor{black}{8.5 to 10.5} Gyr before present \textcolor{black}{(see \citealt{2009AJ....138.1243H}, Figure 11)}.}
		
		
		
	\end{deluxetable*} \label{tab:RLM_fits}
	\subsection{Pulsation-Mass-Radius Relations} \label{sec:PMR_relations}
	We now need a pulsation-mass-radius relation to determine the mass of these stars (\emph{e.g.}, \citealt{1982ApJ...259..198F,1986ApJ...311..864O}).
	\citet{2019MNRAS.482..929T} have done an extensive study of pulsation for long period variables, extending over a greater parameter range than our data.
	Their formula for overtone pulsation is very similar to that found in previous works, and the formulas are reproduced in Table \ref{tab:PMR_fits}.
	
\begin{figure*}[h]
		\centering
		\gridline{
			\fig{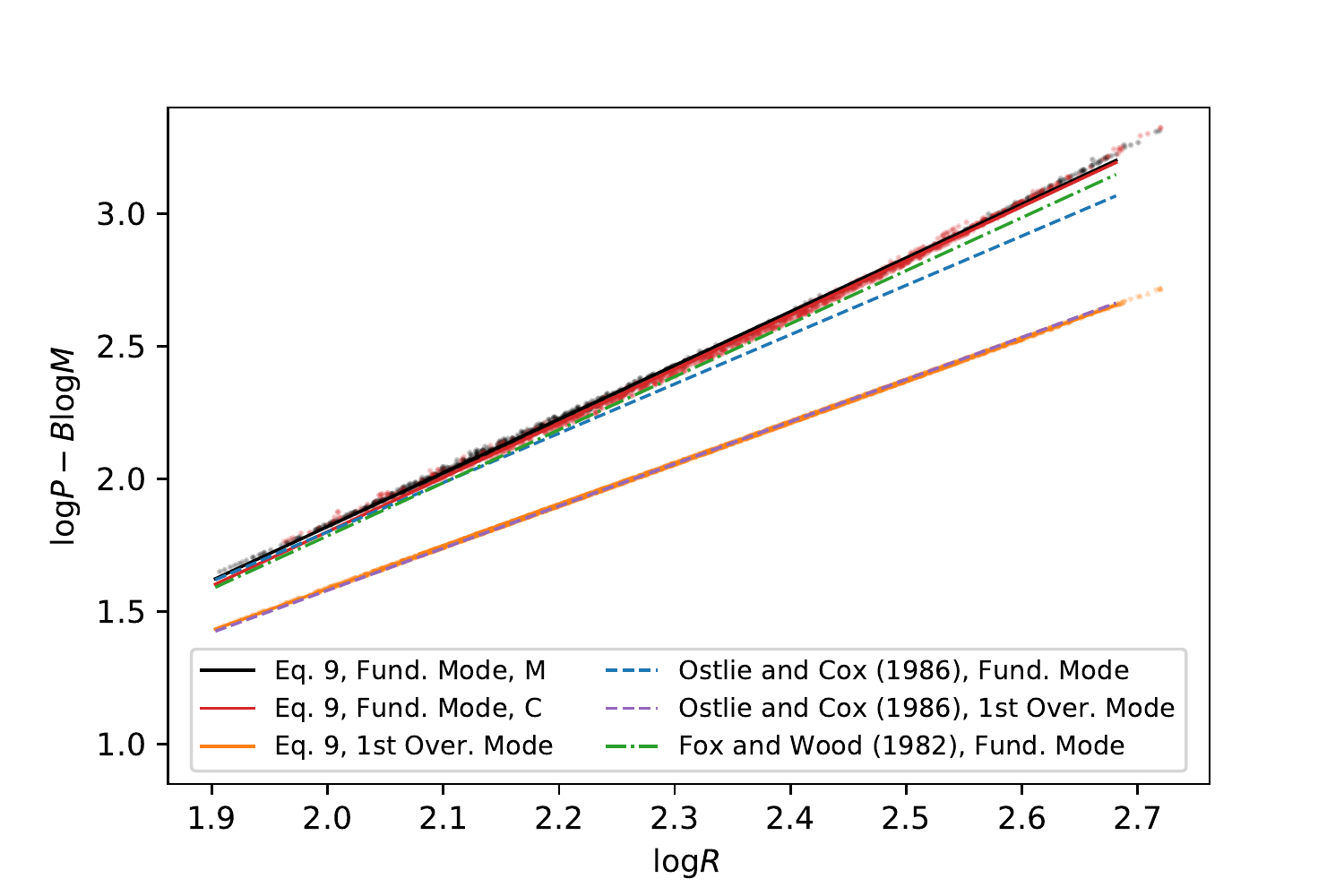}{0.49\textwidth}{(a)} 
			\fig{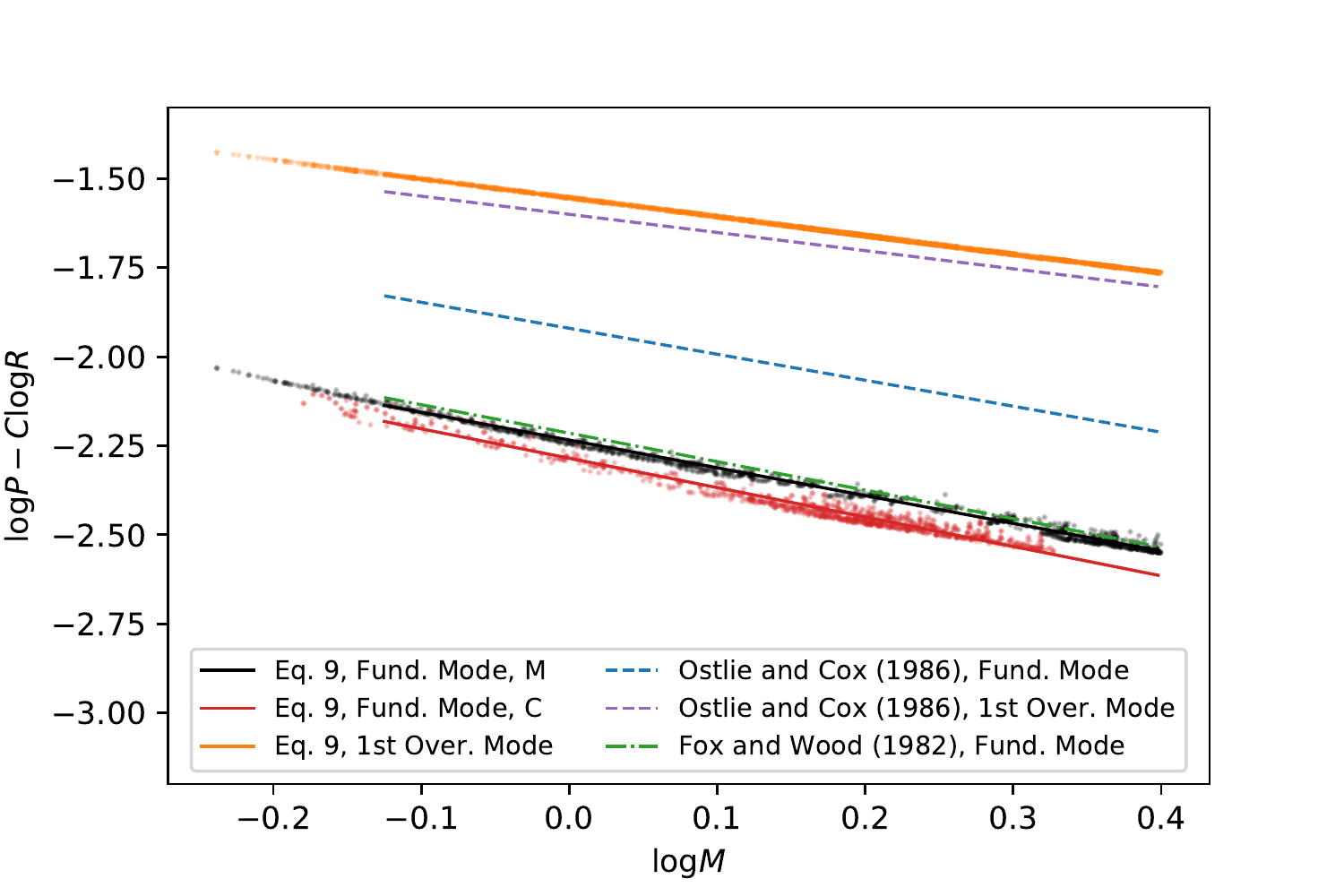}{0.49\textwidth}{(b)} 
		}
		\caption{Comparison of the Pulsation-Mass-Radius (PMR) relations found in this work and \citet{2019MNRAS.482..929T} (solid lines, black for M0, red for C0, and orange for first-overtone) compared to the frequently used relations from \citet{1982ApJ...259..198F} (dotted-dashed) and \citet{1986ApJ...311..864O} (dashed; blue for fundamental mode, purple for first-overtone mode). (a) has had the mass dependence of the relations removed and (b) has had the radius dependence removed, to project a three-dimensional fit into two dimensions. The models used to determine the relations are marked as same color dots on each plot. Note that neither the Fox and Wood formula nor the two Ostlie and Cox formulae distinguish M- and C-stars. While this does not seem to have a significant impact on the relation between pulsation period and radius, we can see a distinct offset when comparing pulsation period and mass. The radius dependence of all formulae are in general agreement with each other. The mass dependence of the formulae are generally in agreement, but with different scaling coefficients; the first-overtone formulae have a small difference while our fundamental mode formulae agree much better with the Fox and Wood formula than the Ostlie and Cox formula. }
		\label{fig:PMR_comparisons}
	\end{figure*}

	For the fundamental mode, \citet{2019MNRAS.482..929T} give a non-linear multi-parameter fitting formula:
	\begin{align}
		\log P = &a_{0} + a_{M} \log M + a_{R} \log R + b_{M} (\log M)^2 + b_{M R} \log M \log R + b_{R} (\log R)^{2} \nonumber \\
		&+ c_{M} (\log M)^{3} + c_{MR} (\log M)^2 \log R + c_{R M} \log M (\log R)^2 + c_{R} (\log R)^3 \nonumber \\ 
		&+ a_{Z} \log Z + a_{Y} Y + a_{\mathrm{C/O}} \log \left[\frac{\mathrm{C/O}}{(\mathrm{C/O})_{\mathrm{ref}}}\right]. \label{eq:PMR_Trabucchi_Fundamental}
	\end{align}
	\textcolor{black}{The \citet{2012ApJ...753...71R} data set lacks information on helium content $Y$ and metallicity $Z$, so we must assume a reasonable value for the entire data set;
		as before we \textcolor{black}{took isochrones of $Z=0.003$ which we then algebraically adjusted to $Z=0.001$} and \textcolor{black}{now} $Y=0.28-Z$.
		As explained in \citet{2019MNRAS.482..929T} and \citet{2021MNRAS.500.1575T}, fitting a fundamental-mode pulsation formula over a wide range of parameters is very difficult due to complex dependence on $M$ and $R$ and the relationship is non-linear over the full parameter range ($0.6 < \log P < 3.8$).
		However, as noted in \citet{2021MNRAS.500.1575T} and as we can see in Figure \ref{fig:PMR_comparisons}, a linear relationship is sufficient when \textcolor{black}{$\log R < 2.4$ for the M0 stars (only one M0 star in our data set is larger), and as noted in \citet{2019MNRAS.482..929T} a linear relationship has errors of up to 20\% for the C0 stars}.
	}
	Thus, we have both the fundamental and overtone mode relations in the form:
	\begin{equation}
		\log P = \log A_{P M R} + B_{P M R} \log M + C_{P M R} \log R \textcolor{black}{+D_{P M R} \log \frac{Z}{0.003} + E_{P M R}(0.003-Z)}, \label{eq:pmr_relation}
	\end{equation}
	\textcolor{black}{noting that this relation is calibrated with the chosen isochrone metallicity $Z=0.003$ and the helium content being $Y=0.28-0.003=0.277$.}
	\textcolor{black}{This relation for first-overtone stars is valid up to 250 days, after which it becomes less precise \citep{2019MNRAS.482..929T}.
		We have applied this relation to 26 of our C1 stars that have periods longer than 250 days, however they make up a small part of the sample. }
	The results for the fundamental modes can be found in Table \ref{tab:PMR_fits}, with the fundamental mode results being in slightly better agreement with the results of \citet{1982ApJ...259..198F} and the overtone mode results being in agreement with \citet{1986ApJ...311..864O}.
	These are good fits to \textcolor{black}{the periods derived from} the \citet{2019MNRAS.482..929T} formula (eq. \ref{eq:pmr_relation}).
	Ideally, this fit would be done with the original models that fell within our range of parameters, because their formula may not be an ideal fit in that subset of parameter space.
	A graphical comparison of the PMR relations described here and those found in \citet{1982ApJ...259..198F} and \citet{1986ApJ...311..864O} can be seen in Figure \ref{fig:PMR_comparisons}.
	\begin{deluxetable}{ccccccc}[hb]
		\tablecaption{Fit of $\log P = \log A_{P M R} + B_{P M R} \log M + C_{P M R} \log R \textcolor{black}{+D_{P M R} \log Z/0.003 + E_{P M R}(0.003-Z)}$}
		\tablehead{\colhead{Subset} & \colhead{$\log A_{P M R}$} & \colhead{$B_{P M R}$} & \colhead{$C_{P M R}$} & \colhead{\textcolor{black}{$D_{P M R}$}} & \colhead{\textcolor{black}{$E_{P M R}$}}} 
		\startdata
		M0 \hfill & -2.234(5)  & -0.7802(2) & 2.027(2) & -0.02713 & 0.14872 \\
		M1 \hfill &  -1.554(4) & -0.529(1)  & 1.570(1) & &\\
		C0 \hfill & -2.285(7)  & -0.826(4)  & 2.043(3) & -0.02713 & 0.14872\\
		C1 \hfill & -1.554(4) & -0.529(1)  & 1.570(1) & &\\
		\enddata
		\tablecomments{$R$ and $M$ are measured in solar units, $P$ is measured in days. 
			\textcolor{black}{Isochrones with metallicity $Z$ were $Z=0.003$, used to find these relations.
				The metallicity adjustments are assumed to be the same as in \citet{2019MNRAS.482..929T}, and are calibrated here for the chosen isochrone metallicity and a helium content $Y=0.28-0.003=0.277$.}
			Values in parentheses are the uncertainty, using standard error.}
	\end{deluxetable} \label{tab:PMR_fits}
	\newpage
	\section{Mass-loss Rates From Stellar Parameters} \label{sec:mdot_from_parameters}
	The ultimate goal of this work is to find reliable formulae for mass-loss rates for AGB stars in terms of stellar parameters $(L, P)$ or $(L, M)$, making use of the extensive data on AGB stars in the LMC.
	The first widely used formula, from \citet{1975psae.book..229R}, expressed mass-loss rate as a power law of luminosity, mass, and radius: $\dot{M} \sim LR/M$.
	Other power law formulations followed:  \citet{1979QJRAS..20..361G,1995AandA...297..727B,2002AandA...384..452W}. 
	\textcolor{black}{\citet{1993ApJ...413..641V}} fitted an exponential dependence on period $P$.
	The formulae based on observations tend to be less steep \textemdash \ that is, to have smaller exponents \textemdash \ than those found from mass-loss modeling (e.g. \citet{1988ApJ...329..299B, 1991ApJ...375L..53B,2000A&A...361..641W,2008PhST..133a4008W,2009ASPC..412..255W,2019A&A...623A.119B,2019A&A...626A.100B}).
	\textcolor{black}{We will compare our results with these earlier formulae in Section \ref{sec:comparison_with_other_formulae}.}
	
	Our standard bilinear fits, described in \textcolor{black}{S}ection \ref{sec:mdot_as_f_L_P}, overestimate low mass-loss rates and underestimate high-mass-loss rates.
	This is characteristic of most of the other published formulae, as we will see in Section \ref{sec:comparison_with_other_formulae}.
	By fitting with $\log L$ as the dependent variable, we show that this is consistent with regression dilution \textcolor{black}{\citep{10.2307/1412159,1999_10.2307/2680496,carroll2006measurement}}, a phenomenon that occurs when the scatter in one or more of the independent variables (in this case, $L$) are too large.
	
	In \textcolor{black}{s}ection \ref{sec:mdot_as_f_L_M}, we have developed an independent method for deriving the exponents from the distribution of the stars in the $\log L$, \textcolor{black}{$\log P$}-plane.
	This method directly relates the distribution of stars in this plane to the exponents of a power law mass-loss formula, and by construction maintains a constant quality-of-fit.
	This method yields significantly larger exponents than a simple linear fit to $\log \dot{M} (\log L, \log P)$ while providing a relation with far less spread than a fit to $\log L (\log \dot{M}, \log P)$. 
	In all cases, the data set has been separated into the five mode and spectral-class based categories established in Section \ref{sec:photometric_catalog}.
	
	\subsection{Mass-loss rate found via multi-linear regression}\label{sec:mdot_as_f_L_P}
	
\begin{figure*}[h]
		\centering
		\gridline{
			\fig{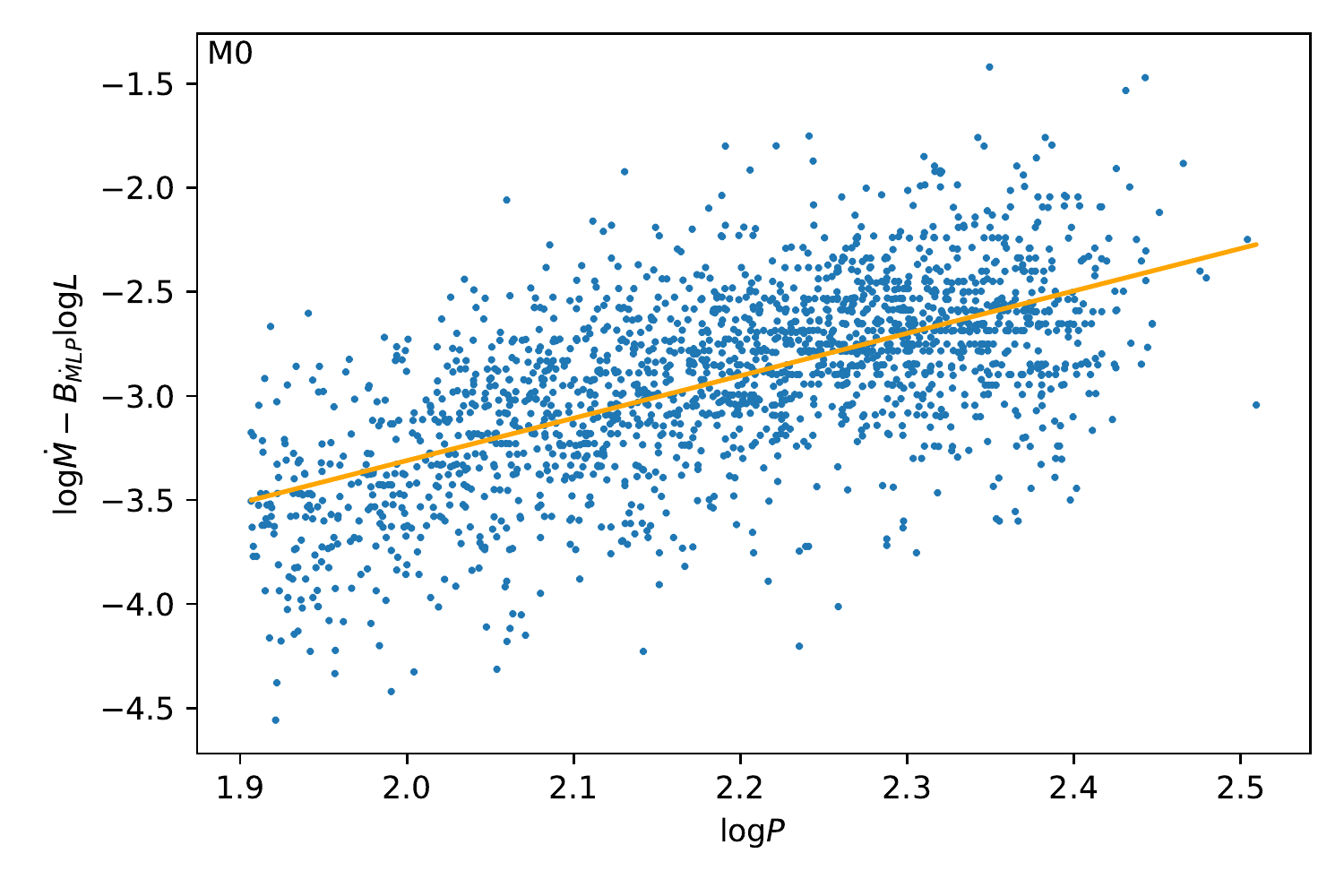}{0.49\textwidth}{(a)} 
			\fig{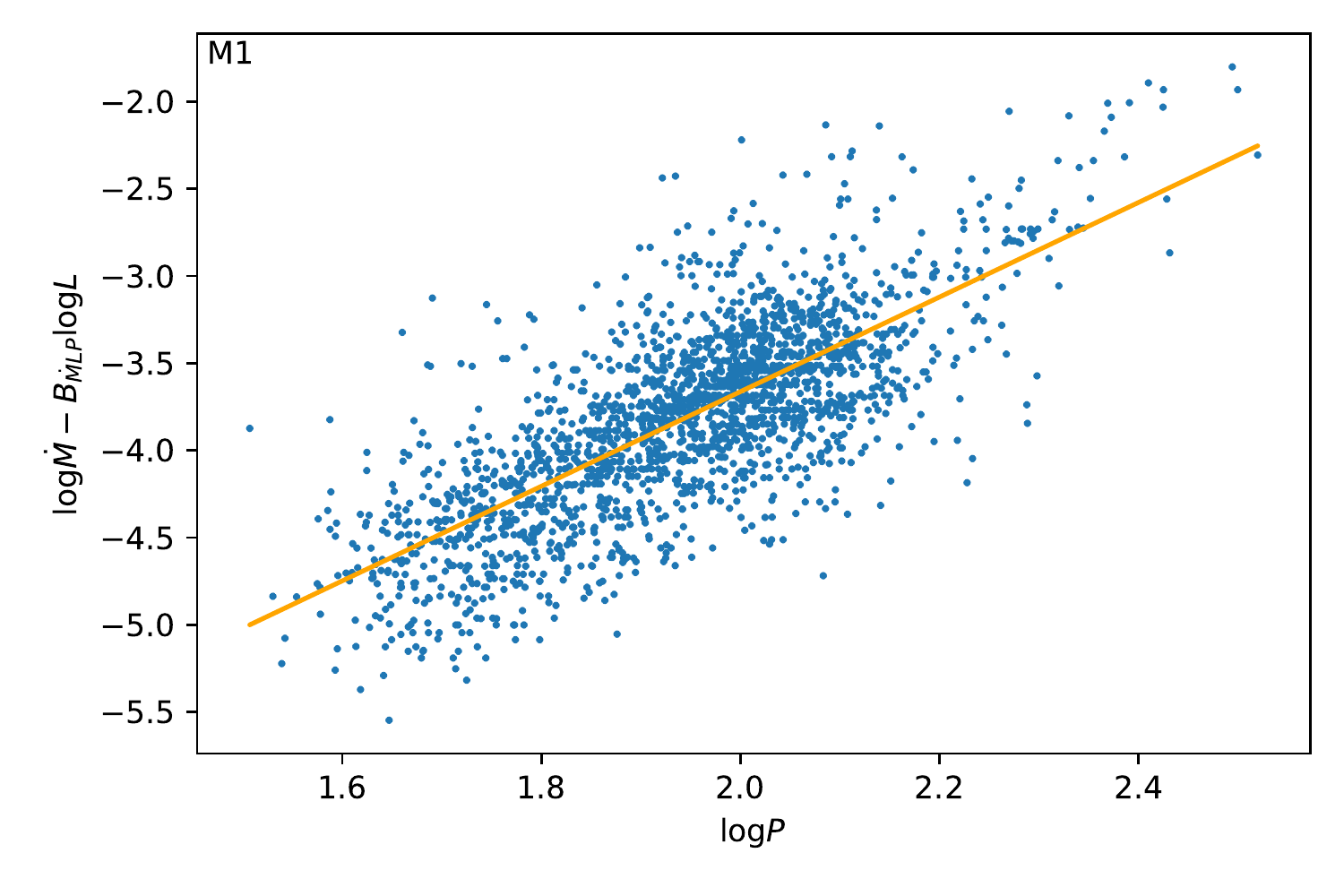}{0.49\textwidth}{(b)} 
		}
		\gridline{
			\fig{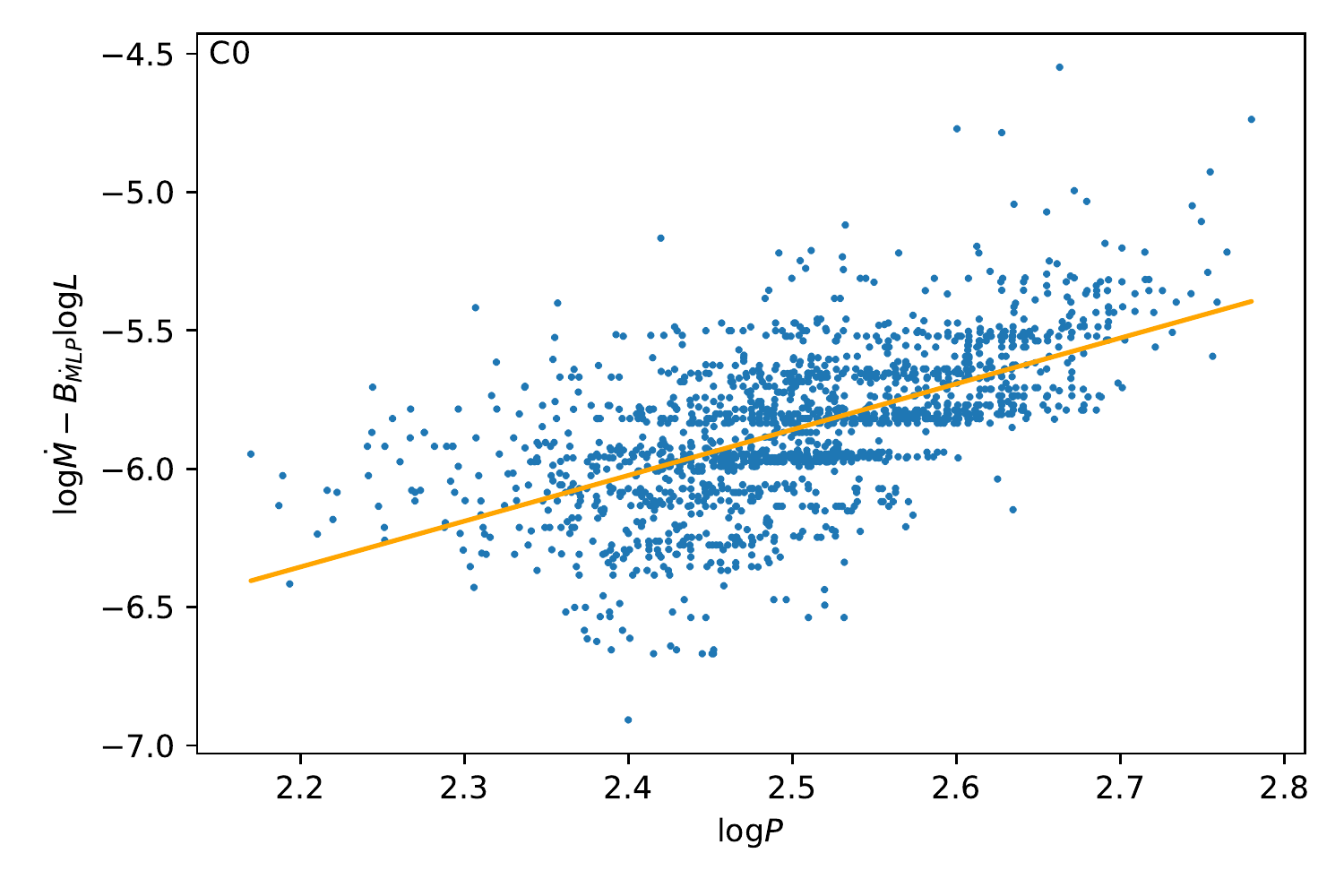}{0.49\textwidth}{(c)} 
			\fig{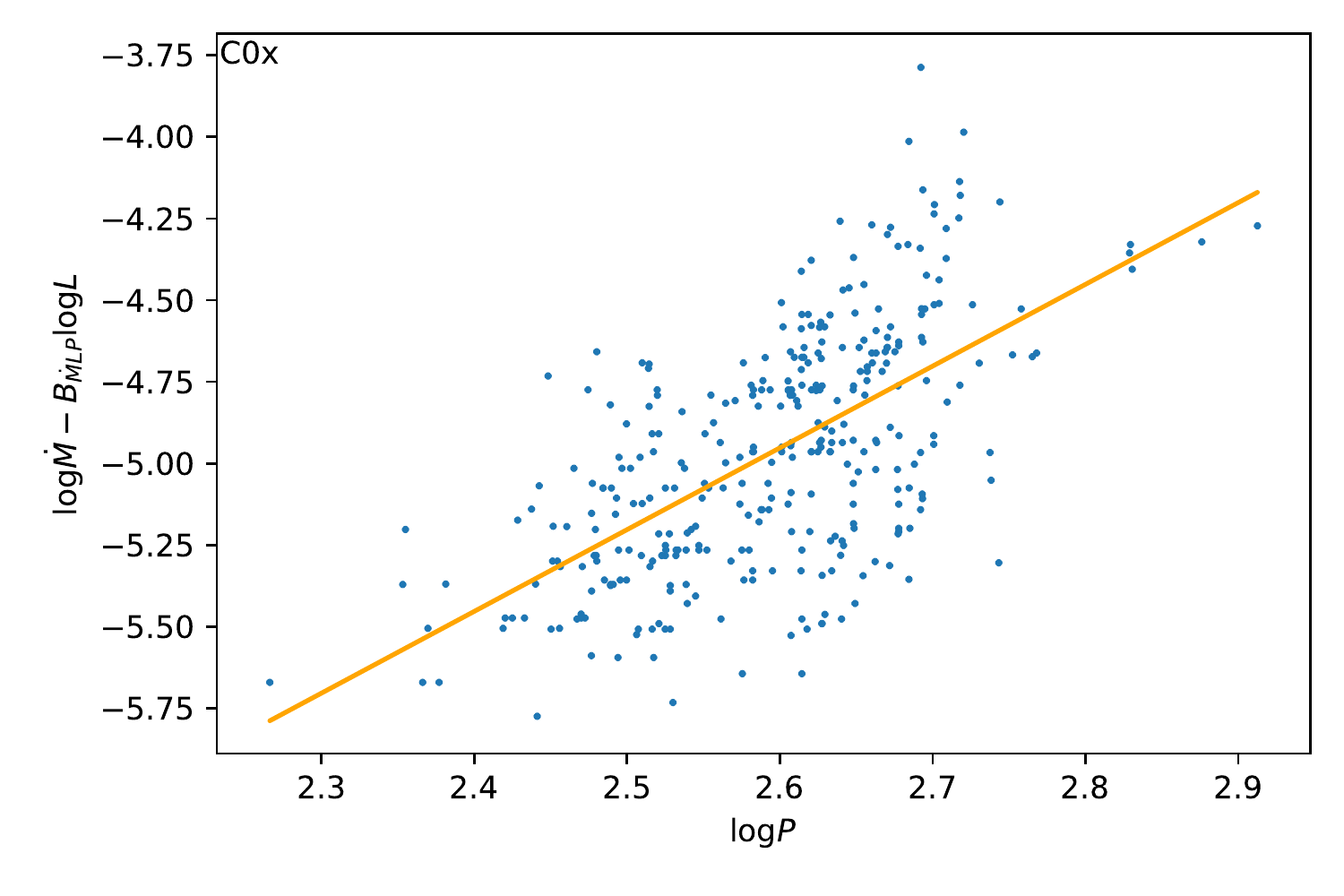}{0.49\textwidth}{(d)} 
		}
		\gridline{
			\fig{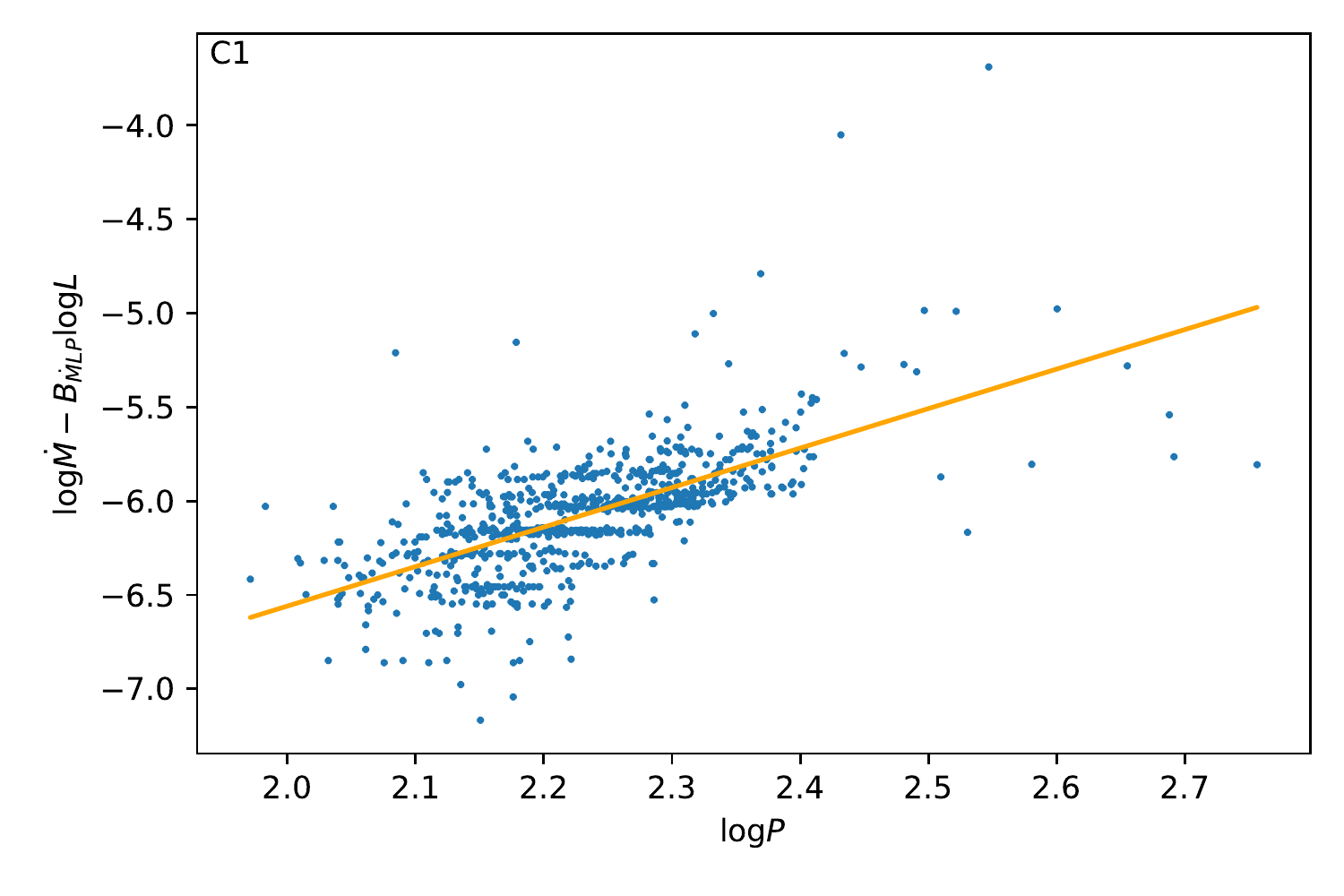}{0.49\textwidth}{(e)} 
		}
		\caption{The quality of the fit of our mass-loss rate formula $\log \dot{M} = \log A_{\dot{M} L P} + B_{\dot{M} L P}\log L + C_{\dot{M} L P}\log P$ from equation \ref{eq:logMdot_LP_fit} for our five combinations of spectral class and pulsation mode. 
			On the $x$-axis is $\log P$ and on the $y$-axis is $\log \dot{M} - B_{\dot{M} L P}\log L$.
			The full details of each fit can be found in Table \ref{tab:logMdot_GRAMS_vs_logL_logP}. Despite the least-squares fit working as intended, we will see in Section \ref{sec:comparison_with_other_formulae} that these fits are experiencing regression dilution.  }
		\label{fig:logMdot-BlogL_vs_logP}
	\end{figure*}

	First, we used a multi-linear least-squares regression to obtain mass-loss rate formulae for AGB stars, using the method built into the Python ``statsmodels'' package \citep{seabold2010statsmodels} and fitting to the equation:
	\begin{align}
		\log \dot{M} = \log A_{\dot{M} L P} + B_{\dot{M} L P}\log L + C_{\dot{M} L P}\log P. \label{eq:logMdot_LP_fit}
	\end{align}
	In this fit and continuing forward, $\dot{M}$ is the total mass-loss rate.
	It was calculated using the dust-to-gas ratio determined using equations \ref{eq:van_loon_dust_to_gas_M} and \ref{eq:van_loon_dust_to_gas_C}, with the dust mass-loss rate being that found by fitting GRAMS models to the AGB stars in the LMC by \citet{2012ApJ...753...71R}.
	We discuss in more detail why these values were chosen in Section \ref{sec:photometric_catalog}.
	Graphical depictions of these fits can be found in Figure \ref{fig:logMdot-BlogL_vs_logP}.
	
	A reliable fit should be reproducible when we switch our dependent and independent variables; otherwise, we will need to look to other methods for verification.
	\textcolor{black}{The refined MACHO measurements have a frequency-space error of $0.00003\ \mathrm{day}^{-1}$ \citep{2008AJ....136.1242F}.
		This corresponds to pulsation period errors of between 0.09\% and 2.5\% for the stars in our selection, with a median error of 0.49\%, \emph{i.e.} are relatively small.}
	\textcolor{black}{We note that individual period errors may be significantly larger, especially in the case of sparse data; verifying these periods is beyond the scope of this paper.}
	However, this is not the case \textcolor{black}{for} luminosity or mass-loss rate.
	We first use luminosity as our dependent variable, and fit the equation:
	\begin{align}
		\log L &= \log \alpha_{L \dot{M} P} + \beta_{L \dot{M} P} \log \dot{M} + \gamma_{L \dot{M} P} \log P.
	\end{align}
	We then algebraically solve for $\dot{M}$ to have the same form as equation \ref{eq:logMdot_LP_fit}, a function of the best fit coefficients:
	\begin{align}
		\log \dot{M} &= \log A_{L \dot{M} P} + B_{L \dot{M} P} \log L + C_{L \dot{M} P} \log P,
	\end{align}
	where $\log A_{L \dot{M} P} \equiv -\log \alpha_{L \dot{M} P}/\beta_{L \dot{M} P}$, $B_{L \dot{M} P} \equiv 1/\beta_{L \dot{M} P}$, and $C_{L \dot{M} P} \equiv -\gamma_{L \dot{M} P}/\beta_{L \dot{M} P}$.
	This procedure results in drastically different values for our exponents.
	This suggests there is a significant amount of regression dilution occurring.
	Regression dilution occurs due to the large uncertainties in the independent variable, here the luminosity as reported by \citet{2012ApJ...753...71R}.
	The results of these fits for our five populations of stars\textemdash M0, M1, C0, C0x, and C1\textemdash can be found in Table \ref{tab:logMdot_GRAMS_vs_logL_logP}.
	A graphical comparison of the quality of these two linear fits and the method discussed in \textcolor{black}{S}ection \ref{sec:mdot_as_f_L_M} can be found in Figure \ref{fig:other_formulae_calc_vs_obsv}.
	
	\textcolor{black}{The natural next step would be to attempt to correct for the regression dilution in some standard way.
		However, there are several reasons why that approach does not work in this case.
		Overall, correction is complicated by the data having differential error.
		The first method attempted was an orthogonal regression, but this method is only valid in cases of multiple independent variables.
		This problem has multiple interdependent variables ($L$, $P$, $M$, and $R$) all of which are connected by the relations found in Section \ref{sec:rm_model_grids}.
		The angular bisector of the two linear fits also fails as a solution because it forces exponents to be between those found in the fits\textemdash as we will see in Sections \ref{sec:mdot_as_f_L_M} and \ref{sec:comparison_with_other_formulae}, better solutions are outside this range.
		Finally, data correction methods for the set fail due to the current state of the problem\textemdash we are attempting to determine the known relation here and we lack covariances for the errors \citep{carroll2006measurement}.  }
	\begin{deluxetable*}{ccccccc}[h]
		
		
		
		
		\tablecaption{Fit of $\log {\dot{M}} = \log A + B\log L + C\log P$}
		
		
		\tablehead{\colhead{Subset} & \colhead{$\log A_{\dot{M} L P}$} & \colhead{$B_{\dot{M} L P}$} & \colhead{$C_{\dot{M} L P}$} & \colhead{$\log A_{L \dot{M} P}$} & \colhead{$B_{L \dot{M} P}$} & \colhead{$C_{L \dot{M} P}$} } 
		
		\startdata
		M0 &  -7.4(2) &  -1.01(7) &  2.04(9) & 8(1) & -12.3(9) & 12(1) \\
		M1 &  -9.1(1) &  -0.77(9) &  2.7(1) & 12(2) & -24(3) & 35(4) \\
		C0 & -10.0(1) & -0.28(5) & 1.66(8) & 4(2) & -15(3) & 18(3) \\
		C0x & -1.0(3) & -0.04(2) & 1.75(8) & -25(14) & -24(12) & 42(21)  \\
		C0, C0x & -11.4(2) & -1.16(6) & 3.6(1) & -4.1(8) & -7.3(4) & 10.0(6) \\
		C1 & -10.8(2) & -0.22(8) & 2.1(1) & 24(10) & -22(8) & 25(9) \\
		\enddata
		
		
		\tablecomments{\textcolor{black}{Comparison of results of fitting $\dot{M}(L, P)$ and translating a fit of $L(\dot{M},P)$.} $\dot{M}$ is measured in solar masses per year, $L$ is measured in solar units, and $P$ is measured in days. $\dot{M}$ is taken to be the dust mass-loss rate multiplied by the inverse of the dust-to-gas ratio $\psi$ (see \textcolor{black}{S}ection \ref{sec:photometric_catalog}). Values in parentheses are the uncertainty, using standard error. As C0x appear to be the C0 stars undergoing the highest mass-loss rates, they are also analyzed together.}
		
		
	\end{deluxetable*}
	\label{tab:logMdot_GRAMS_vs_logL_logP}
	\newpage
	\subsection{Mass-loss Rate found from the \textcolor{black}{PL} strip}\label{sec:mdot_as_f_L_M}
	
	We can use the distribution of AGB stars in the \textcolor{black}{$\log P$, $\log L$} plane as an alternative way to estimate the exponents of a mass-loss formula in those two variables. 
	The general behavior of AGB stars in this plane is described in Figure 2 of \citet{1991ApJ...375L..53B}, and Figure 7 of \citet{2000ARA&A..38..573W}. 
	At the beginning of the AGB these stars present low mass-loss rates barely affecting their total mass, while their core grows due to shell burning, causing their luminosity to increase. 
	In this phase, the stars move horizontally in the diagram (roughly constant $M$ with increasing $L$). 
	This initial trend eventually comes to an end, as finally the mass-loss rate increases to a point that the total mass of the star is significantly affected on timescales shorter than the luminosity increase driven by the slower nuclear timescales. 
	In this phase, the stars move roughly vertically in the $\log L$, $\log M$ plane (roughly constant $L$ with rapidly \textcolor{black}{decreasing} $M$). 
	The occurrence of thermal pulses during the TP-AGB phase, as well as the dynamical processes of dust formation in the circumstellar envelope, introduce scatter to these idealized trajectories, as luminosity and mass-loss rate can change significantly over the short timescales of the He-shell burning.
	The longer and more stable inter-pulse quiescent phases are however well characterized by this general behavior.
	
	A sample of AGB stars selected on the basis of having a reliable determination of their mass-loss rate (such as our LMC samples) will spread on the $L$, $M$ plane along the trajectories described above. 
	Their distribution, however, will be limited to a relatively narrow strip (LM strip hereafter) at the interface between the low and high mass loss phases in their AGB evolution. 
	The region above the strip will be sparsely populated because their lower mass-loss rate will be below the minimum threshold set for a reliable determination of this parameter. 
	The region below the strip will also be depleted because stars in this area evolve too quickly out of the AGB to be detected in any significant number. 
	\textcolor{black}{When we plot $P$ vs. $L$ the stars form a similar strip (PL strip hereafter) where stars progress upwards and to the right as luminosity and pulsation period increase over time.
		In this arrangement, stars below the strip have mass-loss rates that are hard to measure while stars above are evolving off the AGB quickly.}
	
\begin{figure*}[h]
		\centering
		\gridline{
			\fig{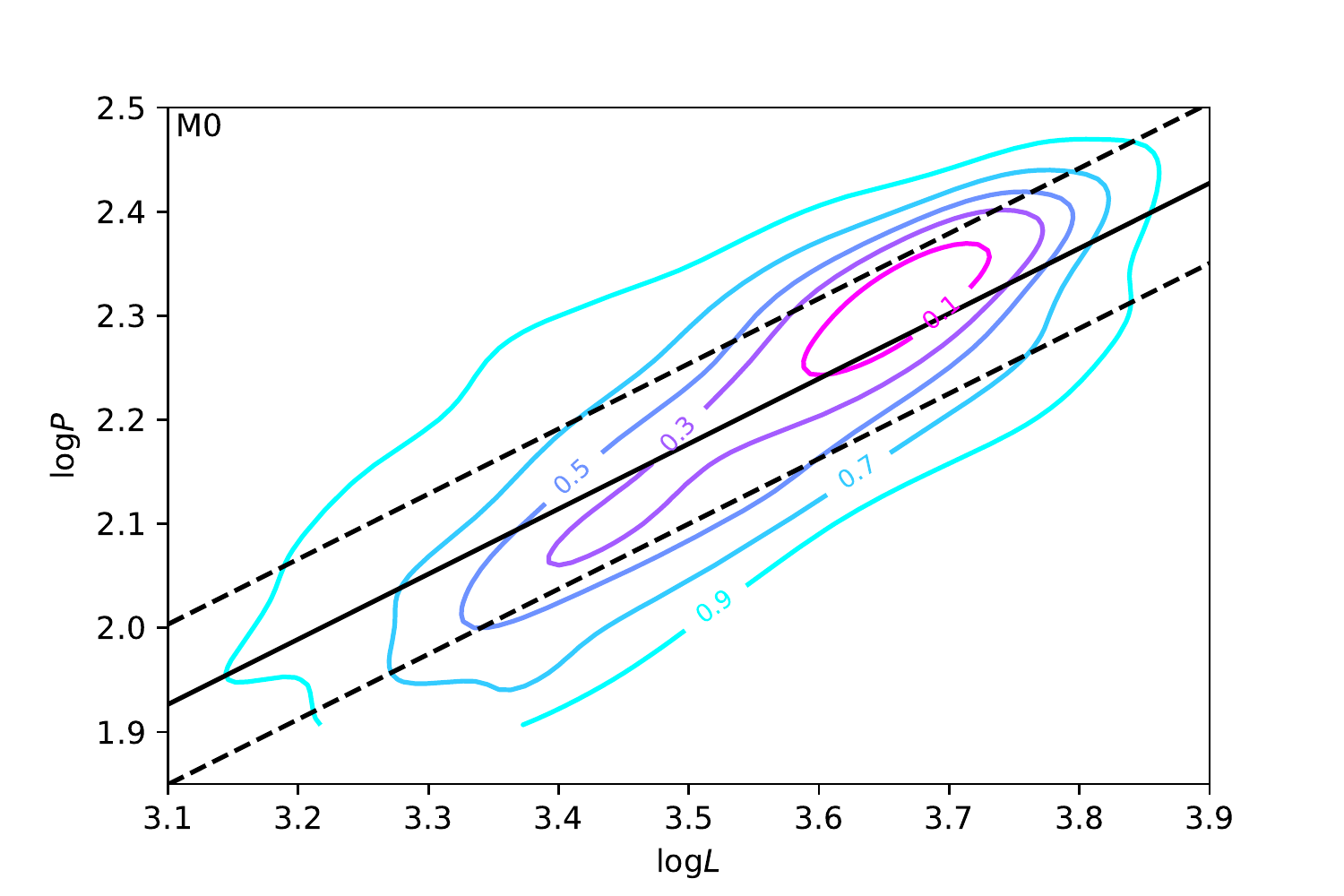}{0.48\textwidth}{(a)}
			\fig{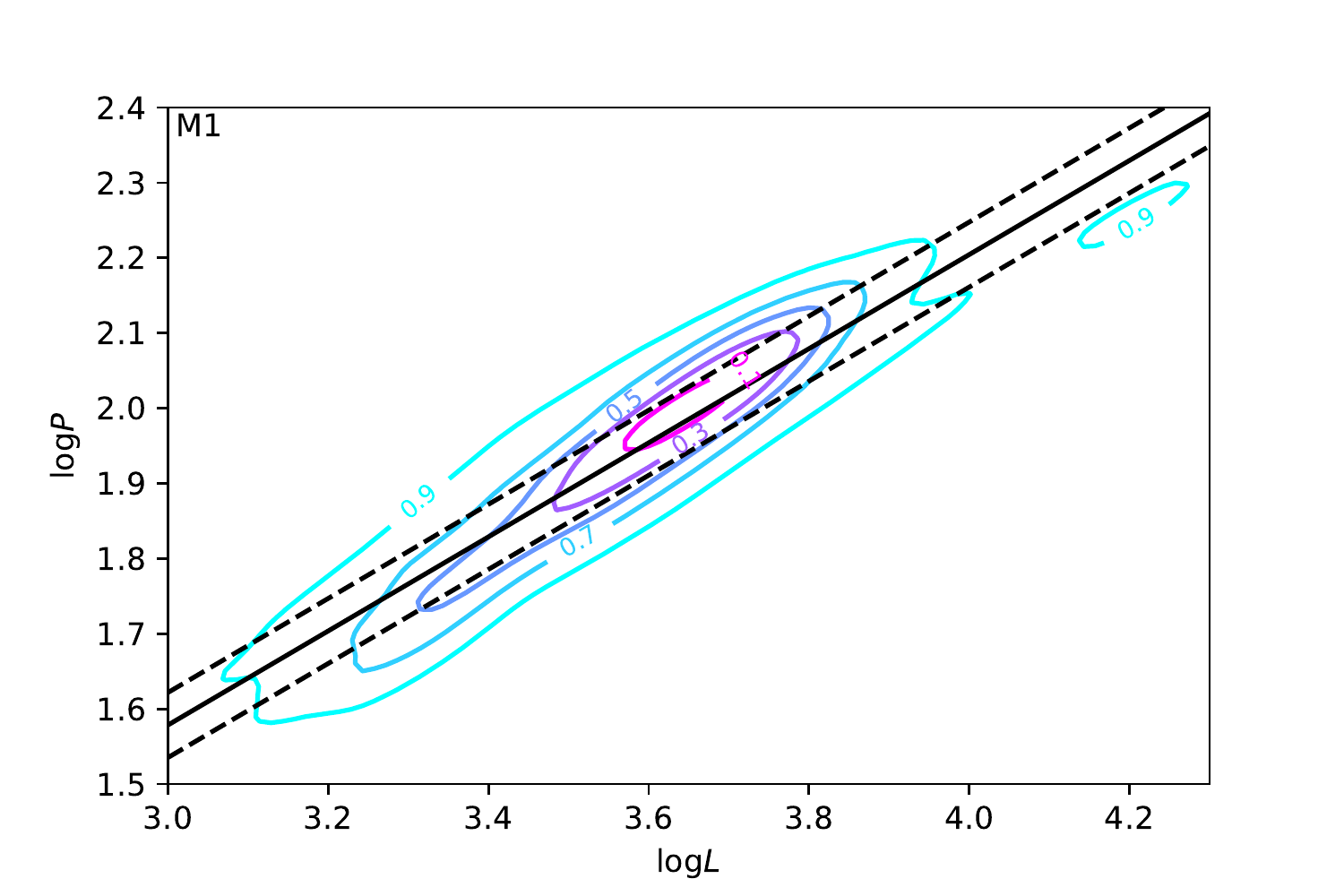}{0.48\textwidth}{(b)}
		}
		\gridline{
			\fig{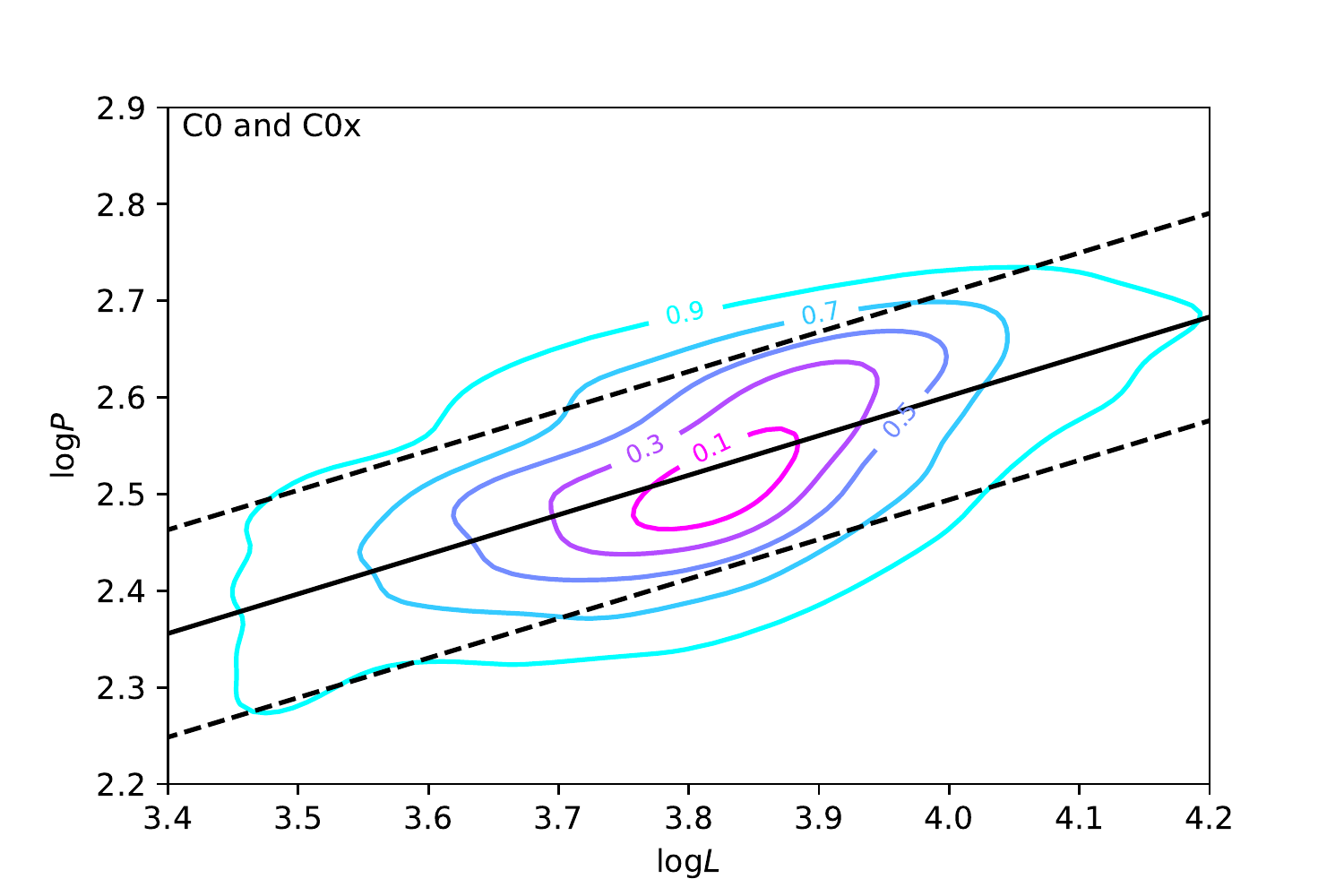}{0.48\textwidth}{(c)}
			\fig{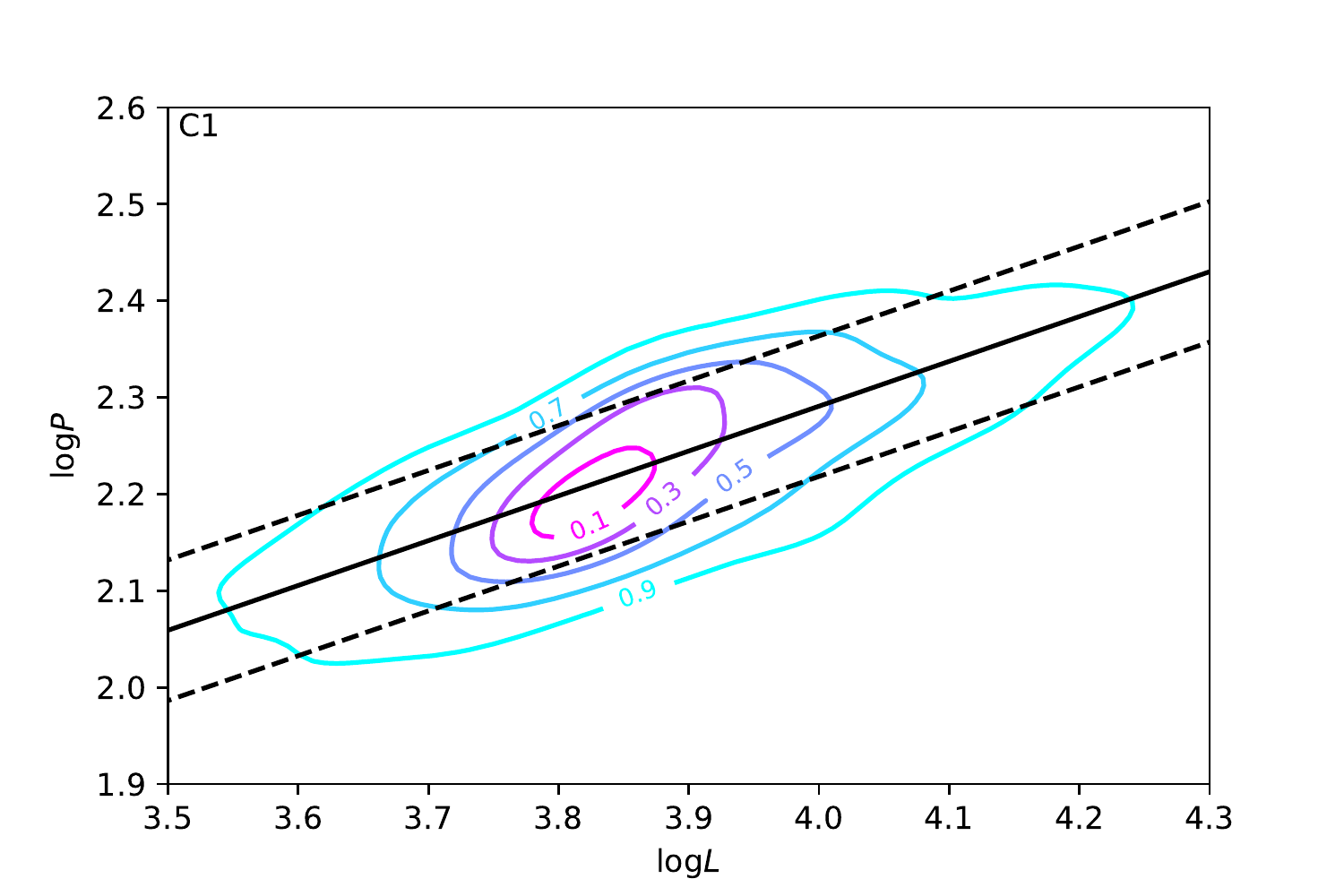}{0.48\textwidth}{(d)}
		}
		\caption{\textcolor{black}{Stars in the PL strip in the \citet{2012ApJ...753...71R} data set, split into five spectral type and pulsation mode categories as indicated on the figures and as detailed in Section \ref{sec:mdot_as_f_L_M}. The black solid line is the best-fit line of each data set, and the black dashed lines bound the stars in the best-fitting strip, with their width and height given in Table \ref{tab:deathzone_fitting_results}. Contours depict \textcolor{black}{the fraction} of stars relative to the peak density. }}
		\label{fig:deathzone_figures}
	\end{figure*}

\begin{figure*}[h]
		\centering
		\gridline{
			\fig{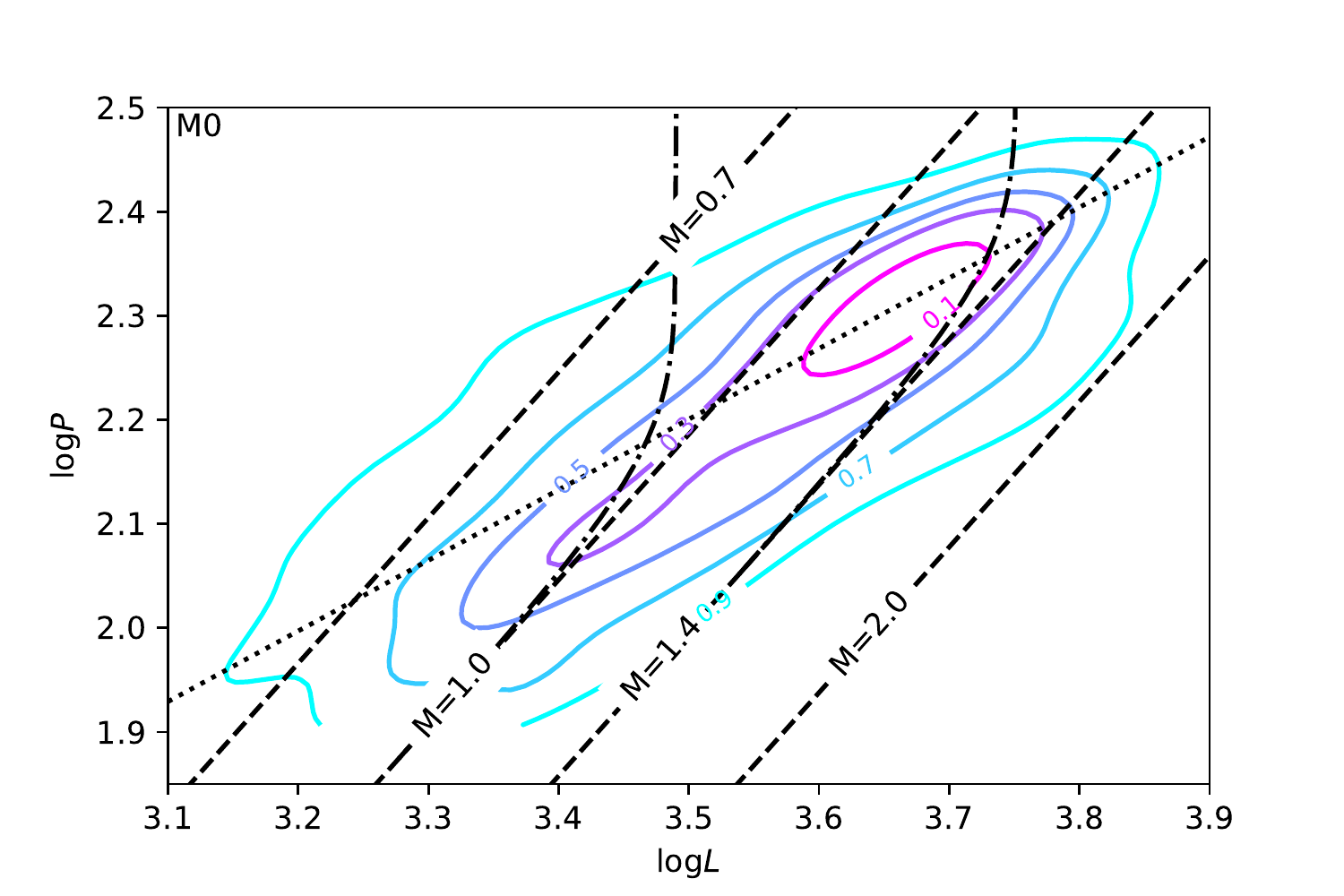}{0.48\textwidth}{(a)}
			\fig{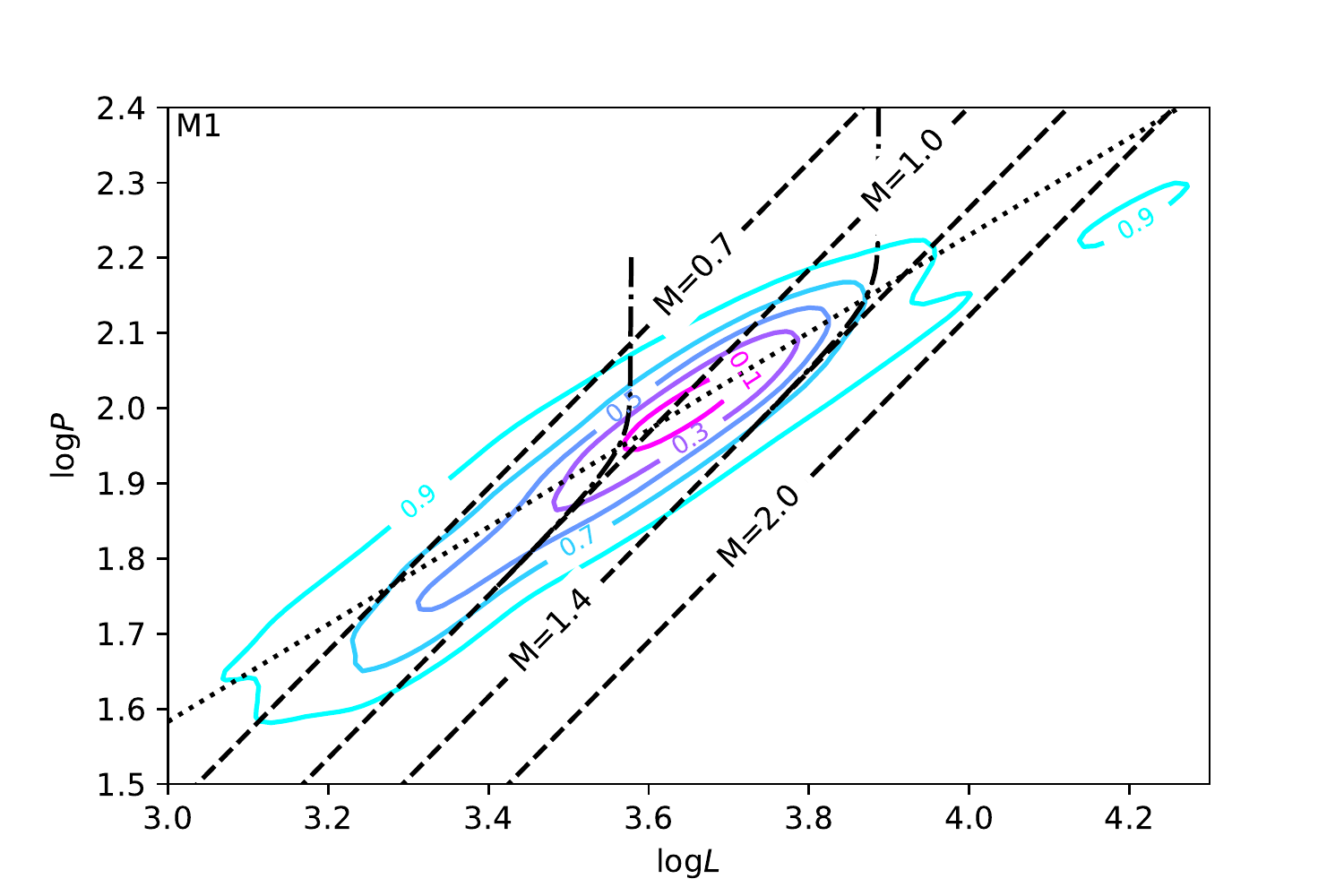}{0.48\textwidth}{(b)}
		}
		\gridline{
			\fig{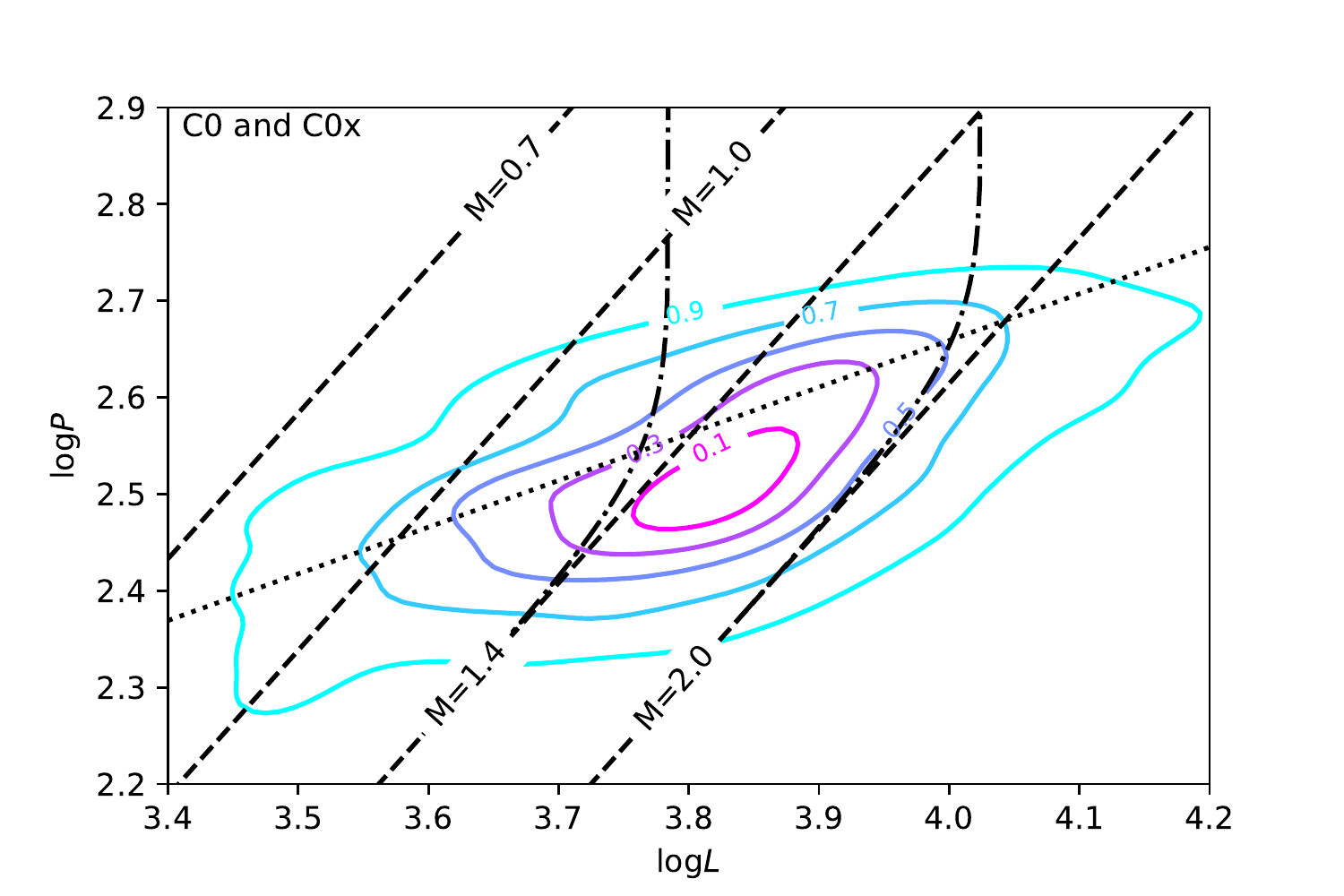}{0.48\textwidth}{(c)}
			\fig{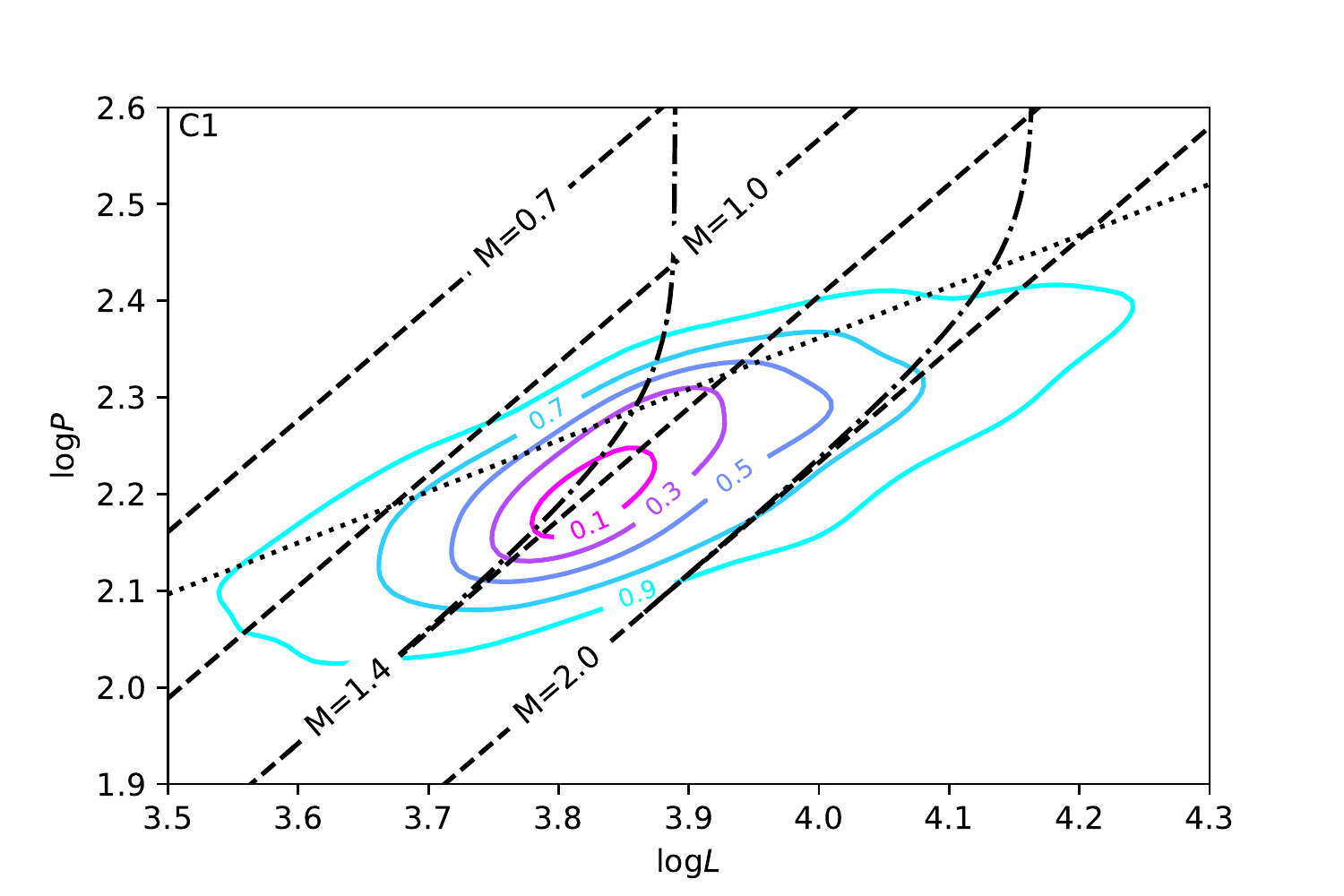}{0.48\textwidth}{(d)}
		}
		\caption{\textcolor{black}{The evolution of the stars using the derived mass loss formulae (Eq. \ref{eq:logMdot_L_P}, Table \ref{tab:deathzone_fitting_results}) is superimposed on the distribution (contours) with the death line indicated (dotted line). Lines of constant mass (dashed lines) and evolutionary tracks (dot-dashed lines) are also included. The pattern is clear: Before the death line, the mass changes relatively little; after, it is seen to be decreasing. The horizontal axis is also a time axis; $\Delta t = (\Delta \log L/2.3)*t_{\mathrm{ev}}$, so with $t_{\mathrm{ev}}=3.2\ \mathrm{Myr}$, $\Delta \log L=0.1$ corresponds to $0.14\ \mathrm{Myr}$.}}\label{fig:deathzone_figures_plus_evo}
	\end{figure*}

	\textcolor{black}{
		With the relations described in Section \ref{sec:rm_model_grids}, we have a pair of linear transformations, which means we will obtain the same result whether we perform the analysis in $\log L$, $\log M$ or $\log L$, $\log P$.
		We are choosing to perform this analysis in $\log L$, $\log P$ because this method uses the untransformed observational data, and so the resulting $\log L$, $\log M$ formulaes can be updated as our understanding of the RML and PMR relations improves.
	}
	This is shown in Figure \ref{fig:deathzone_figures} and \ref{fig:deathzone_figures_plus_evo} for our samples of LMC AGB stars with different envelope chemistry and pulsation modes.
	
	The black solid line shown in Figure \ref{fig:deathzone_figures}, which we derived by fitting the $\log L$ and \textcolor{black}{$\log P$} data in each strip, traces the location where the star’s behavior changes from the low to high mass loss phases described above. 
	This fit line is a good approximation of the ``the cliff'' described in \citet{1991ApJ...375L..53B}, \textcolor{black}{as can be seen in Figure \ref{fig:deathzone_figures_plus_evo}}. 
	
	\textcolor{black}{The conclusion that a majority of the stars are in the death zone }\textcolor{black}{only depends on the distribution of the observed mass-loss rates in $\log \dot{M}/M$ and the chosen evolution time or critical mass-loss rate. 
		As argued in Section \ref{sec:intro}, $t_{\mathrm{ev}}$ should range between \textcolor{black}{$1.2$} and \textcolor{black}{$1.7$} Myr, which does not push a majority of stars out of the death zone at either extreme. 
		The extremes of the dust-to-gas ratio (see equations \ref{eq:van_loon_dust_to_gas_M} and \ref{eq:van_loon_dust_to_gas_C}) also allow bulk shifts of $\textcolor{black}{\log (}\dot{M}/M\textcolor{black}{)}$ by $-0.65$ to $0.33$, which is also insufficient \textcolor{black}{move them outside the death zone}.
		It does not depend on the derived mass-loss formulae, the derived stellar masses and radii, or even the observed periods.
	}
	
	Mass-loss rates increase with increasing $L$ or decreasing $M$, leading to a narrow strip in LM space. 
	The transformation to PL is linear, thus we also have a narrow strip in PL space.
	If the mass-loss rate increases monotonically as $L$ increases or as $M$ decreases (as implied by one- or two-parameter mass-loss rate formulae), then the width of the LM strip depends on how steeply the mass-loss rate increases as a star evolves along the AGB.
	Therefore, for a power law mass-loss formula, larger exponents imply narrower distributions.
	It follows then that by measuring the width and height of the band the stars occupy in $\log M$ versus $\log L$ space we can derive a lower limit on the size of the exponents.
	Absent an overly-restrictive definition of the band, this is a lower limit because any random errors in the measurements will tend to broaden the distribution, so the error-free distribution is as narrow as or narrower than what we find from the observational data.
	This behavior also means that only stars in this region have a well-defined mass-loss law.
	Outside the strip, mass-loss is either too low to effectively be tied with certainty to any of our variables or is in the end phase of AGB evolution where behavior is much more dependent on unobserved variables.
	\textcolor{black}{As long as we have a unique relation between $L$, $M$ and $L$, $P$ we can perform the same analysis in $P$ vs. $L$. }
	
	Our general power law formula has the form:
	\textcolor{black}{
		\begin{equation}
			\log \dot{M} = \log A_{\dot{M} L P} + B_{\dot{M} L P} \log L + C_{\dot{M} L P} \log P.\label{eq:logMdot_L_P}
		\end{equation} 
	}
	By taking the partial derivatives with respect to $\log \dot{M}$, we can show:
	\textcolor{black}{
		\begin{align}
			B_{\dot{M} L P} = & \frac{\partial \log \dot{M}}{\partial \log L} \approx \frac{\Delta \log \dot{M}}{\Delta \log L} \\
			C_{\dot{M} L P} = & \frac{\partial \log \dot{M}}{\partial \log P} \approx \frac{\Delta \log \dot{M}}{\Delta \log P}
		\end{align}
	}
	where $\Delta \log L$ and \textcolor{black}{$\Delta \log P$} are the ranges of luminosity and mass when keeping the other fixed \textemdash\ thus, the width and height of \textcolor{black}{PL} strip, respectively.
	For each class of stars, the observations yield a range of mass-loss rates: $\Delta \log \dot{M}$.
	
	To apply this, we need to precisely define the strip as well as measure the range of mass-loss rates of stars in the strip, excluding any major outliers.
	We will first find the line that best fits the \textcolor{black}{$\log P$} and $\log L$ data.
	\textcolor{black}{
		\begin{equation}
			\log P = \alpha \log L + \beta \label{eq:LM_strip_fit_line}
		\end{equation}
	}
	The \textcolor{black}{PL} strip is defined to be bound in $\log L$-\textcolor{black}{$\log P$} space by lines $\pm n$ \textcolor{black}{times the} standard deviation $\sigma_{FL}$ \textcolor{black}{of the points around} the best fit line.
	\textcolor{black}{This makes} $n$ our fitting variable.
	The width $\Delta \log L$ and height \textcolor{black}{$\Delta \log P$} can then be calculated algebraically using the lines bounding the strip.
	\textcolor{black}{
		\begin{align}
			&\Delta \log P \equiv 2n \sigma_{FL} \cos \left( \tan^{-1} \alpha \right) \\
			&\Delta \log L \equiv \frac{\Delta \log P}{\alpha} 
		\end{align}
	}
	This fixes the ratio of $B$ and $C$ to $\alpha$, the slope of the strip's best-fit line.
	We estimate the range of $\dot{M}$ as the 95\% of stars closest to the mean mass-loss rate; this accounts for a majority of the range while cutting the most extreme outliers.
	This gives $\Delta \log \dot{M} \approx 1.7$ for each set except for C1, where it is about $1.3$; exact values can be found in Table \ref{tab:deathzone_fitting_results}.
	$\Delta \log \dot{M}$ does not change predictably with $n$, so we must determine $n$ by testing different values; for each subset, we checked $0.5 \leq n \leq 3.0$, in steps of $0.0001$.
	The final value of $n$ for each subset stars is whichever value brings the linear fit of $\log \dot{M}_{\textup{calc.}}$ vs $\log \dot{M}_{\textup{obsv.}}$ closest to a slope of 1.
	The scaling coefficient $\log A$ can be determined by forcing the mean residual of $\log \dot{M}$ to be zero.
	
	The results of this method can be found in Table \ref{tab:deathzone_fitting_results}.
	Comparing these results to those we found in Section \ref{sec:mdot_as_f_L_P}, we can see that the \textcolor{black}{PL} Strip method produces large exponents like the indirect linear fit of $\log L(\log \dot{M}, \log P)$, in contrast to the smaller exponents of the direct fit of $\log \dot{M} (\log L, \log P)$. 
	This is further discussed in \textcolor{black}{S}ection \ref{sec:comparison_with_other_formulae} and in Table \ref{tab:residual_slopes_and_rsq} within.
	\begin{deluxetable}{ccccccccccc}[h]
		
		
		
		
		\tablecaption{\textcolor{black}{PL} Strip Analysis of \citet{2012ApJ...753...71R} Data Set}
		
		
		\tablehead{\colhead{Sample} & \colhead{$\alpha$} & \colhead{$\beta$} & \colhead{$n_{\textup{fit}}$} & \colhead{Stars in Strip} & \colhead{$\Delta \log P$} & \colhead{$\Delta \log L$} & \colhead{$\Delta \log \dot{M}$} & \colhead{$\log A_{\dot{M} L P}$} & \colhead{$\log B_{\dot{M} L P}$} & \colhead{$\log C_{\dot{M} L P}$}}
		\startdata
		M0        & \textcolor{black}{0.63} & \textcolor{black}{-0.01} & 1.06 & 1,237 & \textcolor{black}{0.15} & \textcolor{black}{0.24} & 1.62 & \textcolor{black}{-6.63} & \textcolor{black}{-6.67} & \textcolor{black}{10.7} \\
		M1        & \textcolor{black}{0.63} & \textcolor{black}{-0.30} & 0.89 & 1,207 & \textcolor{black}{0.09} & \textcolor{black}{0.14} & 1.92 & \textcolor{black}{-0.26} & \textcolor{black}{-13.9} & \textcolor{black}{22.2} \\
		C0 \& C0x & \textcolor{black}{0.41} & \textcolor{black}{0.96} & 1.51 & 1,668 & \textcolor{black}{0.21} & \textcolor{black}{0.52} & 1.88 & \textcolor{black}{-14.7} & \textcolor{black}{-3.59} & \textcolor{black}{8.77} \\
		C1        & \textcolor{black}{0.46} & \textcolor{black}{0.44} & \textcolor{black}{1.16} & \textcolor{black}{619} & \textcolor{black}{0.15} & \textcolor{black}{0.31} & \textcolor{black}{1.24} & \textcolor{black}{-10.1}  & \textcolor{black}{-3.94}  &  \textcolor{black}{8.49}
		\enddata
		
		
		\tablecomments{$L$ and $M$ are measured in solar units. Exponents $B$ and $C$ are associated with Equation \ref{eq:logMdot_L_P}. As C0x appear to be the C0 stars undergoing the highest mass-loss, they are analyzed together. For further detail, see Section \ref{sec:mdot_as_f_L_M}.}
		
	\end{deluxetable} \label{tab:deathzone_fitting_results}
	
	
	\newpage
	\hfill
	\newpage
	\hfill
	\newpage
	\textcolor{black}{
		\section{Comparison with Other Formulae} \label{sec:comparison_with_other_formulae}
	}
	In this paper, we have approached the problem of finding an expression for the mass-loss rate as a function of stellar parameters in several different ways, each one applied to five sets of stars. 
	We have found rough agreement between two methods, fitting $L(\dot{M},L)$ (\textcolor{black}{S}ection \ref{sec:mdot_as_f_L_P}), and the $L,M$ distribution method (\textcolor{black}{S}ection \ref{sec:mdot_as_f_L_M}), and an explanation for why direct fitting of $\dot{M}$ as a function of stellar parameters produces a discordant result: regression dilution when one independent parameter has too much scatter. 
	In this section, we look at a variety of published formulae, and show that they also show the signature of regression dilution, confirming our conclusion that errors in L have systematically led to underestimates for the exponents in power law fits for the mass-loss rate. 
	
	These formulae do not exist in isolation, so we should examine them in the context of other formulae and other known results.
	Below, you can find a non-exhaustive list of other observation-based formulae from over the last 45 years.
	\begin{align}
		\text{\citet{1975psae.book..229R}:\ } \dot{M} = &\left(4 \times 10^{-13} \right) \eta \frac{L R}{M} \label{eq:Reimers1975}\\
		\text{\citet{1977A&A....61..217R}:\ } \dot{M} = &\left(4 \times 10^{-13} \right) (0.35) \frac{L R}{M}\label{eq:Reimers1977} \\
		\text{\citet{1993ApJ...413..641V}:\ } \log \dot{M} = &\begin{cases}
			-11.4 + 0.0123P & \text{ if } M<2.5\ \mathrm{M_{\odot}} \\ 
			-11.4 + 0.0125(P-100(M-2.5)) & \text{ if } M>2.5\ \mathrm{M_{\sun}}
		\end{cases} \label{eq:VW1993}\\
		\text{\citet{1995AandA...297..727B}:\ } \dot{M} = & 1.932\times 10^{-21} \frac{L^{3.7} R}{ M^{3.1}} \label{eq:Bloecker1995}  \\
		\text{\citet{2005AandA...438..273V}:\ } \log \dot{M} = &-5.65 + 1.05 \log \left(\frac{L}{10^{5} L_{\sun}}\right) - 5.3 \log \left(\frac{T}{3500\ \mathrm{K}}\right)\label{eq:vanLoon2005} \\
		\text{\citet{2005ApJ...630L..73S}:\ } \dot{M} = &\left(4 \times 10^{-13} \right) \eta \frac{L R}{M} \left(\frac{T}{4000\ \mathrm{K}}\right)^{3.5} \left( 1 + \frac{g_{\sun}}{g_{\star}}\right)\label{eq:SchroederCuntz2005} \\
		\text{\citet{2010AandA...523A..18D}:\ } \log \dot{M} = &\begin{cases}
			-7.37 + 3.42\times10^{-3}(P) & \text{ if } P<850\ \mathrm{d}  \\ 
			-4.46 & \text{ if } P>850\ \mathrm{d}
		\end{cases}\label{eq:deBeck2010}
	\end{align}
	In order to compare the various published formulae with each other and with this work, where possible we have used our evolutionary tracks and period-mass-radius relations, together with the definition of effective temperature, to algebraically transform the formulae to a power law in $L$ and $M$.
	When this is not possible, we approximated exponents using a bilinear regression in $\dot{M}$, $L$, and $M$, where $\dot{M}$ was calculated with the formula in question for the LMC stars. 
	Our fits also use LMC data, and we do not expect the scaling values ($\log A$) to match across data sets of substantially different metallicity.
	These comparisons can be found in Table \ref{tab:other_formulas_logMdot_logL_logM}.
	\begin{deluxetable*}{cccccc}[h]
		\tablecaption{Equivalent or Estimated Fit to Function $\log {\dot{M}} = \log A + B\log L + C\log M$ for Other Formulae}
		\tablehead{\colhead{Reference} & \colhead{$\log A$} & \colhead{$B$} & \colhead{$C$} &\colhead{Slope of Calc. vs. Obsv. Fit} & \colhead{Std. Dev. of Residual} \\ \colhead{} & \colhead{} & \colhead{} & \colhead{} & \colhead{M0, M1, C0 and C0x, C1} & \colhead{M0, M1, C0 and C0x, C1} }
		\startdata
		\citet{1975psae.book..229R} &  -12.79 & 1.72 & -1.31 & 0.14, 0.32, 0.13, 0.30 & 0.38, 0.37, 0.39, 0.24  \\
		\citet{1993ApJ...413..641V} $^{\textup{a}}$ &  -59.0 & 14.3 & -15.1 & \textcolor{black}{M0: 1.07, C0: }1.32  & \textcolor{black}{M0: 0.34, C0: }0.80 \\
		\textcolor{black}{\citet{1995AandA...297..727B}} & \textcolor{black}{$-21.0$} & \textcolor{black}{$4.39$} & \textcolor{black}{$-3.35$} & \textcolor{black}{$0.35$, $0.81$, $0.33$, $0.77$} & \textcolor{black}{$0.51$, $0.54$, $0.44$, $0.32$} \\
		\citet{2005AandA...438..273V} &  5.66 & 1.73 & -0.98 & 0.12, 0.34, 0.123, 0.31 & 0.40, 0.38, 0.41, 0.25 \\
		\citet{2005ApJ...630L..73S} &  -14.7 & 2.27 & -1.81 & 0.22, 0.46, 0.20, 0.47 & 0.39, 0.38, 0.31, 0.24 \\
		\citet{2010AandA...523A..18D} &  -14.2 & 2.06 & -2.04 & 0.15, 0.14, 0.37, 0.31 & 0.37, 0.39, 0.35, 0.24
		\enddata
		\tablecomments{$\dot{M}$ is measured in solar masses per year, $L$ and $M$ are measured in solar units. Exponents are the average of those found using the evolutionary tracks and PMR relations. The second, \textcolor{black}{fifth}, and \textcolor{black}{sixth} formulae were estimated due to not being power laws. 
			\newline
			$^{a}$ \citet{1993ApJ...413..641V} was determined using fundamental-mode stars so is only valid for \textcolor{black}{those subsets} of AGBs. }
		
	\end{deluxetable*}
	\label{tab:other_formulas_logMdot_logL_logM}
	
\begin{figure*}[ht]
		\centering
		\includegraphics[width=1.0\textwidth]{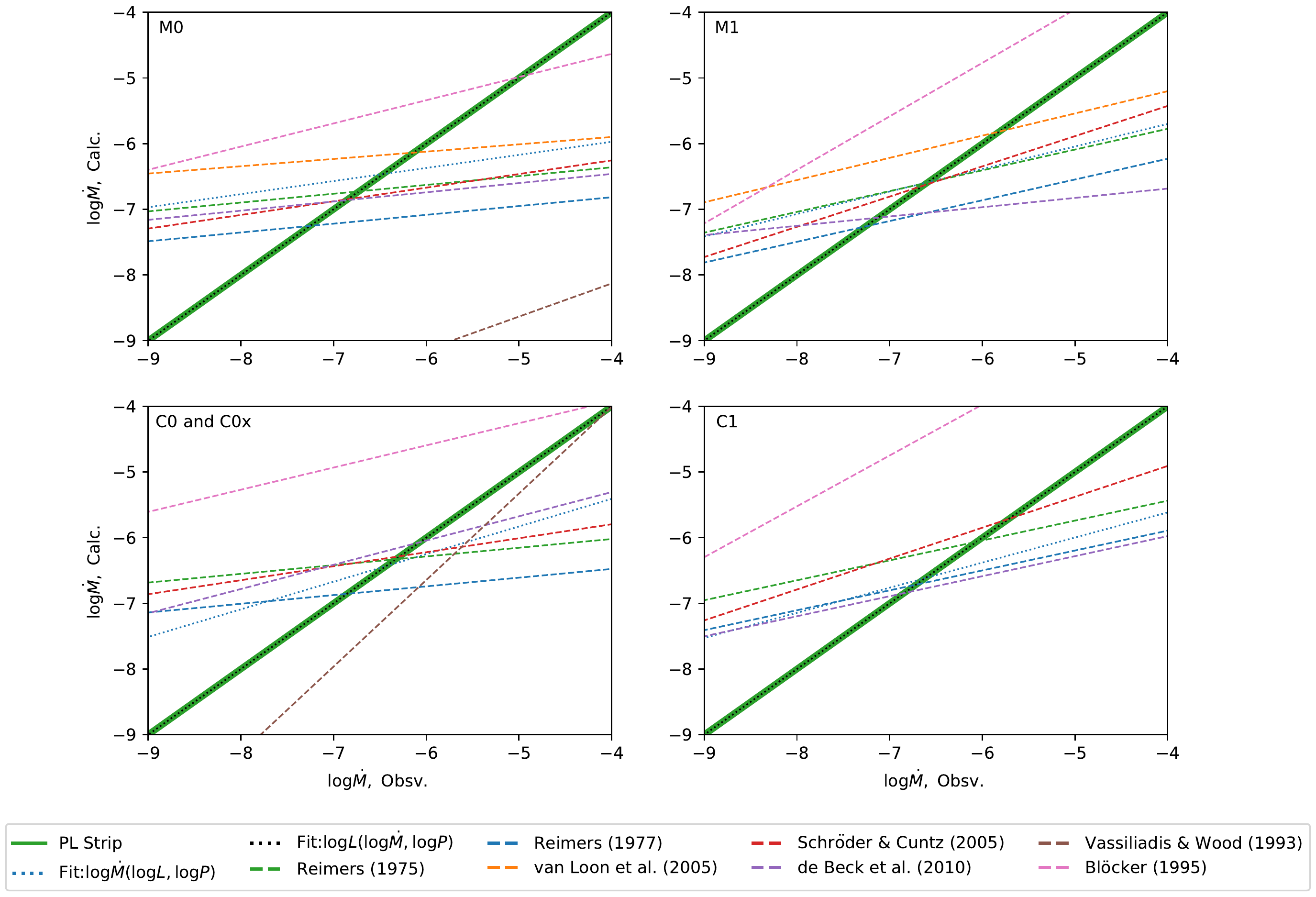}
		\caption{\textcolor{black}{Comparison of the calculated vs. observed mass-loss rates for the formulae we have found and other works.
				The solid line shows our PL strip fit and matches a 1:1 line by construction (Table \ref{tab:deathzone_fitting_results}).
				The dotted lines show the results of the linear fits (Table \ref{tab:logMdot_GRAMS_vs_logL_logP}); the green diagonal line for the PL strip method coincides with what was found by fitting $\log L(\log \dot{M}, \log P)$ (black dotted line).
				The dashed lines show our fits to equations \ref{eq:Reimers1975}-\ref{eq:deBeck2010}.
				The formulae in those equations have been used to calculate a mass-loss rate ($y$-axis), and then a line has been fitted to the result of the formula versus the observed mass-loss rate ($x$-axis).
				\textcolor{black}{Note that the fit lines have been extended beyond the range of data in order to display the formulae in a single plot and to show their divergent behavior.}
				Nearly all the other relations show the low-slope signature of regression dilution, except the formula from \citet{1993ApJ...413..641V}. 
				Note that this formula is offset in the M0 panel, due to a difference in mass-loss rate scaling.
				The spread of data around the various lines can be found in Tables \ref{tab:other_formulas_logMdot_logL_logM} and \ref{tab:residual_slopes_and_rsq}.}
		}
		\label{fig:other_formulae_calc_vs_obsv}
	\end{figure*}

	In Table \ref{tab:residual_slopes_and_rsq}, we can see comparisons of the three formulae we have determined in this paper.
	The multi-linear fit of $\log \dot{M} (\log L, \log P)$ suffers from a notable issue: underestimating high-mass-loss rates and over-estimating low mass-loss rates with a failure to match the one-to-one line in $\log \dot{M}_{\textup{calc.}}$ vs $\log \dot{M}_{\textup{obsv.}}$ space.
	This is the characteristic signature of regression dilution.
	Except for the \citet{1993ApJ...413..641V} formula, which is a fit to the low-noise parameter $P$, the above listed formula (equations \ref{eq:Reimers1975}\textendash \ref{eq:deBeck2010}) all show the same characteristic signature of regression dilution.
	This issue can be seen clearly if $\log \dot{M}_{\mathrm{calc.}}$ is plotted versus $\log \dot{M}_{\mathrm{obsv.}}$, as we have done in Figure \ref{fig:other_formulae_calc_vs_obsv} for the formulae in equations \ref{eq:Reimers1975}-\ref{eq:deBeck2010}.
	This \textcolor{black}{issue is circumvented} in our \textcolor{black}{PL} strip analysis, which finds much better correlation between the calculated and observed values with a similar significance of the fit, as seen by comparing Tables \ref{tab:other_formulas_logMdot_logL_logM} and \ref{tab:residual_slopes_and_rsq}.
	This result strongly implies that linear fitting methods are suffering from regression dilution, due to both the uncertainty in our measurements of $\dot{M}$ and $L$ and the non-independence of the $\dot{M}$ and $L$ measurements, confirming the suggestions of \citet{2005AandA...438..273V, 2006A&A...445.1069G, 2012ApJ...753...71R}; and \citet{2018A&ARv..26....1H}.
	Without knowing the reliability of the measurements, we cannot correct for the bias in the linear fits \textcolor{black}{\citep{10.2307/1412159,1999_10.2307/2680496,carroll2006measurement}}.
	
	\begin{deluxetable*}{cccccccc}
		
		
		
		
		\tablecaption{\textcolor{black}{Determined Formulae in the form} $\log \dot{M} = \log A + B\log L + C \log M$}
		
		
		\tablehead{\colhead{\textcolor{black}{Subset} } & \colhead{Method and} & \colhead{$\log A_{\dot{M} L M}$} & \colhead{$B_{\dot{M} L M}$} & \colhead{$C_{\dot{M} L M}$} & \colhead{$|\frac{B_{\dot{M} L M}}{C_{\dot{M} L M}}|$} & \colhead{Slope of } & \colhead{\textcolor{black}{$\sigma$} } \\ 
			& \colhead{\textcolor{black}{Fitting} Variables} &  &  &  &  & \colhead{Calc. vs. Obsv.} & \colhead{of Residuals} } 
		
		\startdata
		M0 &  Linear Fit - $\log \dot{M}(\log L,\log P)$ &  -12.9 &  1.85 &  -2.63 & 0.70 &  0.21 &  0.35 \\
		M0 &  Linear Fit - $\log L(\log \dot{M},\log P)$ &  -26.4 &  5.72 &  -16.6 & 0.34 & 1.00 &  1.22  \\
		M0 &  \textcolor{black}{PL} Strip \textcolor{black}{Method} &  \textcolor{black}{-35.2} &  \textcolor{black}{8.24} &  \textcolor{black}{-13.7} & 0.59 & 1.00 &  0.80 \\
		\\
		M1 &  Linear Fit - $\log \dot{M}(\log L,\log P)$ &  -14.3 &  2.17 &  -2.50 & 0.87 & 0.34 &  0.36 \\
		M1 &  Linear Fit - $\log L(\log \dot{M},\log P)$ &  -54.8 &  13.5 &  -32.1 & 0.42 & 1.00 &  2.00  \\
		M1 &  \textcolor{black}{PL} Strip \textcolor{black}{Method} &  \textcolor{black}{-42.8} &  \textcolor{black}{10.2} &  -20.5 & 0.50 & 1.00 &  1.19 \\
		\\
		C0 and C0x &  Linear Fit - $\log \dot{M}(\log L,\log P)$ &  -21.9 &  4.23 &  -5.66 & 0.75 & 0.42 &  0.33 \\
		C0 and C0x &  Linear Fit - $\log L(\log \dot{M},\log P)$ &  -33.6 &  7.74 &  -15.8 & 0.49 & 1.00 &  0.82  \\
		C0 and C0x &  \textcolor{black}{PL} Strip \textcolor{black}{Method} &  \textcolor{black}{-41.0} &  \textcolor{black}{9.62} &  -13.9 & 0.69  & 1.00 &  0.52 \\
		\\
		C1 &  Linear Fit - $\log \dot{M}(\log L,\log P)$ &  -15.1 &  2.21 &  -2.34 & 0.94 & 0.38 &  0.23 \\
		C1 &  Linear Fit - $\log L(\log \dot{M},\log P)$ &  -27.4 &  6.47 &  -27.7 & 0.23 & 1.00 &  2.30  \\
		C1 &  \textcolor{black}{PL} Strip \textcolor{black}{Method} &  \textcolor{black}{-27.8} &  5.89 &  -9.44 & 0.62 & 1.00 &  0.51 \\
		\enddata
		
		
		\tablecomments{For ease of comparison \textcolor{black}{with equations \ref{eq:Reimers1975}\textendash \ref{eq:deBeck2010}}, all formulae \textcolor{black}{found in this work} have been transformed into \textcolor{black}{the form $\dot{M} = A L^{B} M^{C}$} using PMR relations and evolutionary tracks of the appropriate kind for spectral class and mode. The \textcolor{black}{PL} Strip provides a fit that first works consistently over the entire range of mass-loss rates while also providing a tighter spread than the multi-linear fit to $\log L$. The \textcolor{black}{PL} Strip formulae should be taken as the correct formulae. }
		
	\end{deluxetable*} \label{tab:residual_slopes_and_rsq}
	\newpage \hfill \newpage
	\textcolor{black}{\section{Results}} \label{sec:results}
	\indent We can see in Table \ref{tab:residual_slopes_and_rsq} and Figure \ref{fig:other_formulae_calc_vs_obsv} that the \textcolor{black}{PL} strip provides the overall best fit, with tolerable spread in the predictions and no change in quality over the range of mass-loss rates.
	These large exponents also tend to agree with the exponential \citet{1993ApJ...413..641V} formula, which agrees well with observations of AGB mass loss found in globular clusters by \citet{2010MNRAS.408..522K} and with results from atmospheric modelling \citep{2000A&A...361..641W,2019A&A...623A.119B,2019A&A...626A.100B}. 
	From atmospheric modelling, we also expect larger exponents than those found in the linear fit to $\dot{M}$, consistent with an abrupt mass-loss phase, but their size is an unresolved question (compare \citet{2000ARA&A..38..573W,2018A&ARv..26....1H}).
	The \textcolor{black}{PL} strip method also has some limits: if the width of the distribution in $\log \dot{M}$ is partly due to observational uncertainty, then correcting for this would make the exponents smaller.
	If the width of the distribution in $\log L$ is partly due to observational error, then correcting for this would make the exponents bigger.
	
	\textcolor{black}{
		The observed pattern of stars in $\log P$ or $\log M$ vs. $\log L$ is what we expect if the death-zone analysis holds, given that the sample only contains stars with measurable mass-loss rates between $10^{-8}$ and $10^{-4}\ \mathrm{M_{\sun}/yr}$.
		This range includes the death line, where $\dot{M} = \dot{M}_{\mathrm{crit}}\equiv M/t_{\mathrm{ev}}$, for any reasonable value of $t_{\mathrm{ev}}$ based on evolutionary models.
		By construction, our power-law exponents produce a strip of the correct width, height, and slope.
	}
	
	\textcolor{black}{
		The position of the death-line relative to the observations is determined by the value of the coefficient $A$ in the power law (eq. \ref{eq:logMdot_LP_fit}) and the value of $t_{\mathrm{ev}} = 1/(\mathrm{d} \ln L/\mathrm{d} t)$.
		The coefficient $A$ depends on the dust-to-gas ratio, and is therefore somewhat uncertain.
		The evolution time is also not trivial to derive, given that many of these stars are experiencing shell flashes.
		However, if our interpretation of the strip is correct, then by following the evolution of the star in $\log M$ vs. $\log L$ (and thus vs. time), we can position the death-line with respect to the location of the bulk of the stars (see Figure \ref{fig:deathzone_figures_plus_evo}).  
	}
	
	\textcolor{black}{In Figures \ref{fig:deathzones} and \ref{fig:mass_histogram}, we can see that whether a star becomes a carbon star is primarily determined by stellar mass.
		Below $1\ \mathrm{M_{\sun}}$, AGB stars are primarily oxygen-rich; above $1.3 \mathrm{M_{\sun}}$ they are primarily carbon-rich.
		Stars between these masses can be of either type.
	}
	
	
	\begin{figure*}[h]
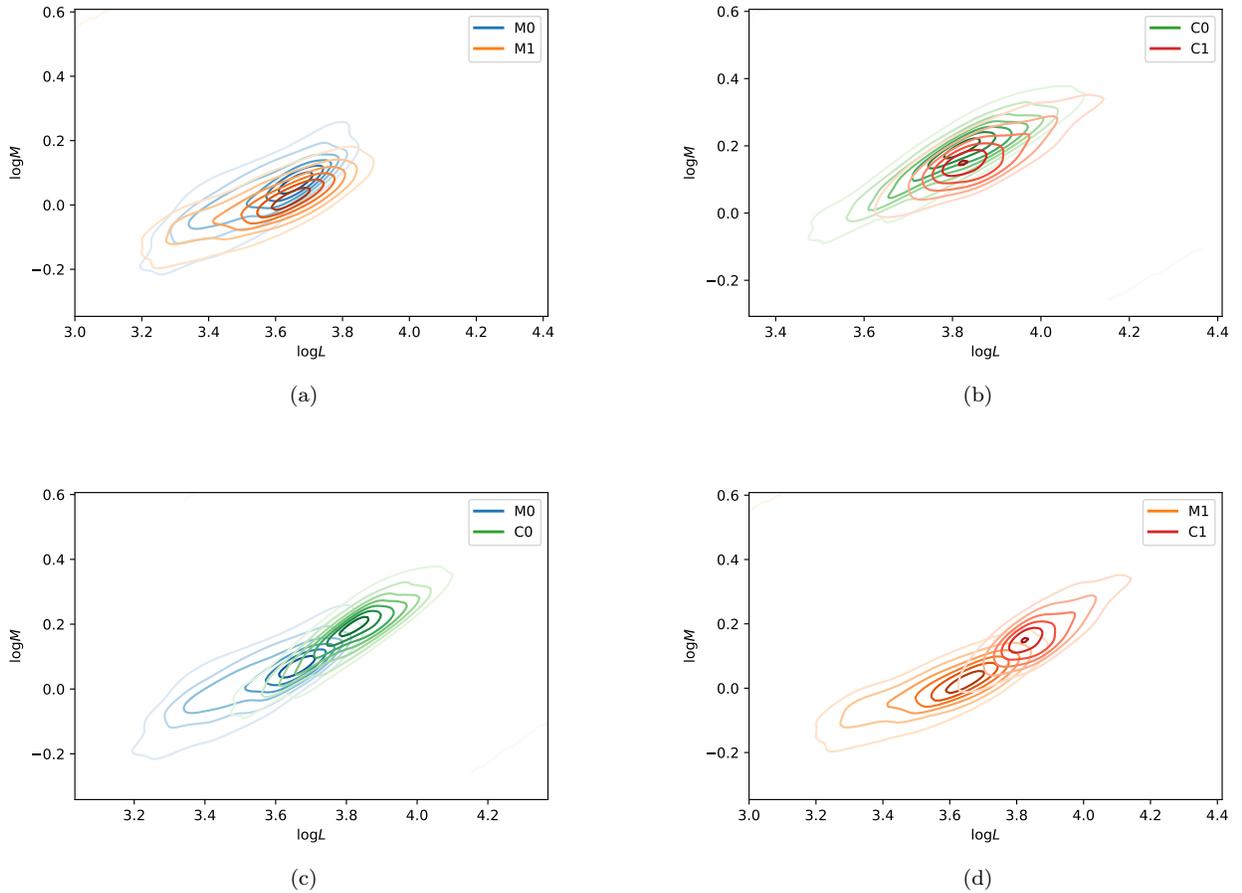

		\gridline{
			\fig{Death_Zone/stnd_dev/LM/LM_contour__M0_M1.pdf}{0.45\textwidth}{(a)} 
			\fig{Death_Zone/stnd_dev/LM/LM_contour__C0_C1.pdf}{0.45\textwidth}{(b)}
		}
		\gridline{
			\fig{Death_Zone/stnd_dev/LM/LM_contour__M0_C0.pdf}{0.45\textwidth}{(c)} 
			\fig{Death_Zone/stnd_dev/LM/LM_contour__M1_C1.pdf}{0.45\textwidth}{(d)}
		}
		\caption{\textcolor{black}{Contour plots depicting the location of the LM strip for the combinations of pulsation mode and \textcolor{black}{C/O} composition. The number of stars in each set can be seen in Table \ref{tab:stellar_stats}. The distribution of stars in (c) and (d) is consistent with our assertion that the death zone is well populated for all four categories of stars.}}
		\label{fig:LM_strip_contour_plots}
	\end{figure*}
	

	\begin{figure*}[h]
		\gridline{
			\fig{Death_Zone/deathlines_t_ev_10^6.5/LM_deathline_M0_M1_C0_C1.pdf}{0.45\textwidth}{(a)}
			\fig{Death_Zone/deathlines_t_ev_10^6.5/LP_deathline_M0_M1_C0_C1.pdf}{0.45\textwidth}{(b)} 
		}
		\caption{\textcolor{black}{Locations of the death lines for our four sets of stars. 
				The death line is defined by the luminosity and pulsation periods of stars where $\dot{M}=\dot{M}_{\textup{crit.}} \equiv M/t_{\textup{ev}}$. The luminosity evolution time scale, $t_{\mathrm{ev}}$, is set to $0.6\ \mathrm{Myr}$ for the M0 stars, $0.5\ \mathrm{Myr}$ for the M1 Stars, and $2.5\ \mathrm{Myr}$ for the C0 Stars, and $3.2\ \mathrm{Myr}$ for the C1 Stars.
				Points are marked every $\Delta \log L = 0.015$, with ``+'' marking fundamental mode points and ``x'' marking overtone mode points; given $\Delta \log L = 2.3 \Delta t/t_{\mathrm{ev}}$, we can see why a majority of stars are expected to be found prior to the death zone.
				The evolution tracks terminate at a final mass determined using the initial-final mass relation in \citet{2012ApJ...746..144Z}.
				The death lines span the tenth to ninetieth percentile in mass for each data set. }
		}
		\label{fig:deathzones}
	\end{figure*}
	
	
	
	\begin{figure*}[h]
		\gridline{
			\fig{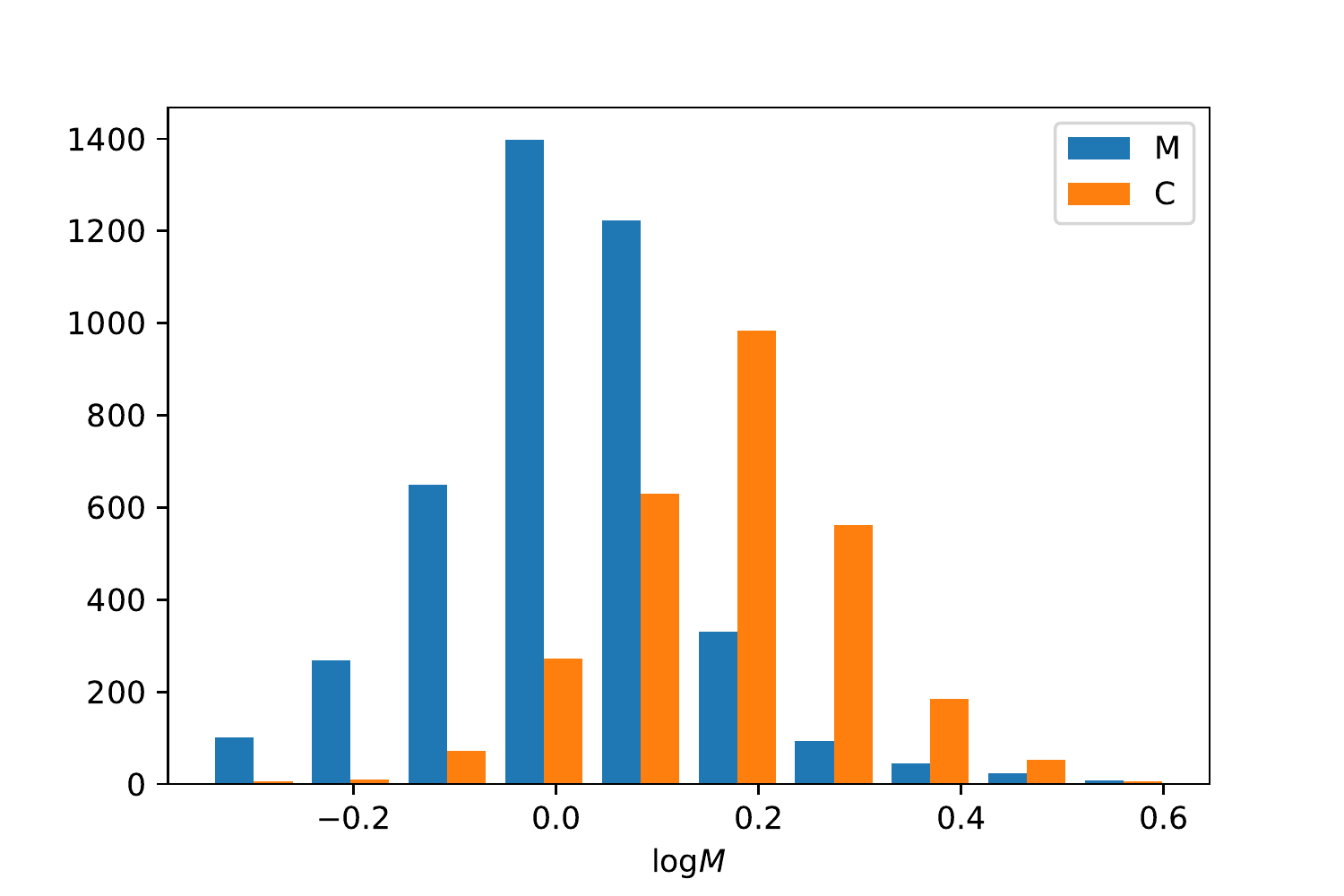}{0.75\textwidth}{}
		}
		\caption{\textcolor{black}{The derived distribution of the masses of the oxygen and carbon-rich stars in our sample, showing that the parameter determining \textcolor{black}{$C/O$} composition is primarily the mass. 
				Mass loss reducing $M$ and shell flashes modulating $L$ introduce scatter in $M$ and may be responsible for most of the overlap in the distributions.}
		}
		\label{fig:mass_histogram}
	\end{figure*}
	\section{Conclusions and Extension}
	We have used a sample of 6,889 LMC AGB pulsating and mass-losing stars \citep{2012ApJ...753...71R} to derive power-law formulae for mass-loss rates as a function of stellar parameters. 
	By approaching the derivation of a mass-loss formula from the observations in three ways, we have discovered why previous approaches have produced very different formulae: uncertainties in the measurement of $L$ produce regression dilution that reduces the exponents when a fit is made to $\log \dot{M}$ as a function of $\log L$ and $\log P$ (or $\log L$ and other derived stellar parameters). 
	Using this insight, and an analysis of the distribution of the stars in \textcolor{black}{$\log P$} versus $\log L$, we have found formulae that satisfy our expectations that the slope of a fit to $\log \dot{M}_{\textup{calc.}}$ versus $\log \dot{M}_{\textup{obsv.}}$ should be equal to one. 
	These \textcolor{black}{PL} strip formulae also reproduce the distribution in the $\log L$, \textcolor{black}{$\log P$} plane, and show relatively small scatter in the residuals. 
	There is a separate formula for each of four groups, two composition groups ($C/O > 1$ and $<1$) and two modes of pulsation, fundamental and first overtone. 
	The exponents in these formulae are closer to what is expected from mass loss models. 
	We conclude that the \textcolor{black}{PL} strip method provides the best formulae that can be derived from this set of observations.
	
	The method we have used can be applied to other samples with sufficient numbers of stars at a known distance, such as the Small Magellanic Cloud or Andromeda.
	
	In the process of deriving these formulae we have used published evolutionary models to derive new relations for the period as a function of mass and radius. 
	This allows us to present the formulae in terms of $\log L$ and $\log P$, the most readily observed quantities, or in terms of $\log L$ and $\log M$, appropriate for use with evolutionary models (taking into account that the relations differ according to pulsation mode).
	
	\textcolor{black}{A robust and surprising conclusion of this study is that the death zone is well populated for all four categories of stars \textemdash \ fundamental and overtone pulsators, oxygen- and carbon-rich stars.
		The carbon and oxygen rich stars separate mostly by mass. 
		The overtone pulsators reach their death zones at slightly higher L than the fundamental mode stars. 
		All four categories are being observed as they go over the cliff.}
	
	\begin{acknowledgements}
		This work has made use of the NumPy library \citep{harris2020array}, the SciPy Library \citep{Virtanen_2020}, IPython software package \citep{PER-GRA:2007}, the matplotlib library for publication quality graphics \citep{Hunter:2007}, and the Scikit-learn library \citep{scikit-learn}.
		\textcolor{black}{Funding for this work has been provided by the New Mexico Space Grant Consortium.}
	\end{acknowledgements}
	\newpage 
	\appendix
	\section{Estimating Dust mass-loss rate From Color} \label{sec:mdot_from_color}
	While looking for ways to reduce scatter in our $\log \dot{M} \sim \log L + \log P$ fits (see Section \ref{sec:mdot_from_parameters} for details), we fitted the mass-loss rate to a hyperbolic curve, as in \citet{1996A&A...311..253B, 1998A&A...334..173L, 2009MNRAS.396..918M}; and \citet{2012ApJ...753...71R}.
	The hyperbolic curves found in these works are of the form
	\begin{align}
		\log \dot{M} = \frac{A}{B + (\mathrm{color})} + C,
	\end{align}
	with the color varying depending on the observational bands available.
	
\begin{figure*}
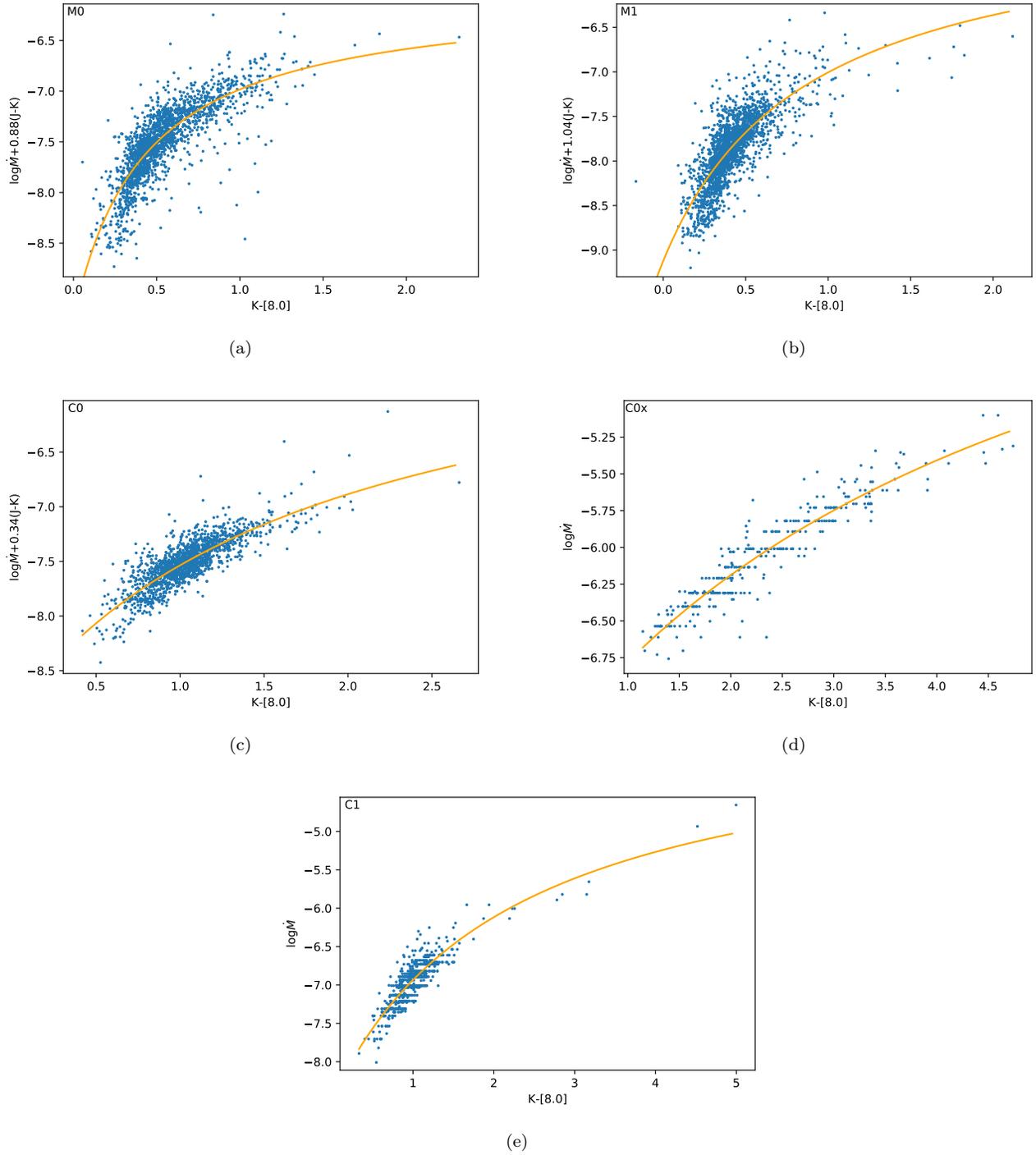

		\gridline{
			\fig{mdot_vs_color_fits/M0logMdot_vs_K-_8.0_.pdf}{0.45\textwidth}{(a)} 
			\fig{mdot_vs_color_fits/M1logMdot_vs_K-_8.0_.pdf}{0.45\textwidth}{(b)}
		}
		\gridline{
			\fig{mdot_vs_color_fits/C0logMdot_vs_K-_8.0_.pdf}{0.45\textwidth}{(c)}
			\fig{mdot_vs_color_fits/C0x_logMdot_vs_K-_8.0_.pdf}{0.45\textwidth}{(d)}
		}
		\gridline{
			\fig{mdot_vs_color_fits/C1_logMdot_vs_K-_8.0_.pdf}{0.45\textwidth}{(e)}
		}
		\caption{The fit of $\dot{M}$ to the $\mathrm{K-[8.0]}$ color, using the function $\log \dot{M} = A/((\mathrm{K-[8.0]})+B) + C - D(\mathrm{J-K})$ is shown. 
			In the case of the oxygen-rich, fundamental-mode pulsators $\mathrm{M0}$, (a); oxygen-rich, first-overtone pulsators $\mathrm{M1}$, (b); and carbon-rich, fundamental-mode pulsators $\mathrm{C0}$, (c), a second color ($\mathrm{J-K}$) was used, and is accounted for by including this dependence on the $y$-axis with $\log \dot{M}$ with the appropriate fitting exponent. 
			For the carbon-rich, extreme, fundamental pulsators $\mathrm{C0x}$, (d) (see Figure \ref{fig:C0_split} for details), and carbon-rich, first-overtone pulsators $\mathrm{C1}$, (e), only a single color was used and thus only $\log \dot{M}$ is found on the $y$-axis. 
			The values of these fits and their accuracy can be found in Table \ref{tab:color_mdot_fits}. }
		\label{fig:logMdot_vs_color_fits}
	\end{figure*}

	Initially, we performed a single-color fit, using $K-[8.0]$ as our primary color as in \citet{2009MNRAS.396..918M} and \citet{2012ApJ...753...71R}.
	We found this fit to be unsatisfactory for our purposes for the M0, M1, and C0 stars, so we continued onto a two-color fit, in $K-[8.0]$ and $J-K$.
	\begin{align}
		\log \dot{M} = \frac{A}{B+ (K-8.0)} + C - D \times (J-K)
	\end{align}
	where $J-K$ is treated as a linear correction to the original hyperbolic fit.
	With this, we were able to obtain fits with significantly better confidences.
	The results of these fits can be found in Table \ref{tab:color_mdot_fits}, with graphical depictions in Figure \ref{fig:logMdot_vs_color_fits}.
	
	Examining Figure \ref{fig:logMdot_vs_color_fits}, it is clear that two colors does not completely predict the mass loss of AGB stars.
	This is unsurprising, as the color is affected directly by factors unique to each star.
	Despite this, the use of near-infrared J and K bands in addition to a mid- to long-wavelengths (such as [8.0]) allows one to make an estimate of the mass-loss rate of an observed star, without the computational burden of fitting to any models.
	
	\begin{deluxetable*}{cccccc}[ht]
		
		
		
		
		\tablecaption{Fit of $\log \dot{M} = A/((K-[8.0]) + B) + C - D\times(J-K)$}
		
		
		\tablehead{\colhead{Spectral Class and Pulsation Mode}  & \colhead{A} & \colhead{B} & \colhead{C} & \colhead{D} } 
		
		\startdata
		M0 &  -1.4(1) &  0.43(5) &  -6.01(8) & 0.88(3) \\
		M1  & -3.4(5)  & 0.87(9)  &  -4.4(2) &  1.74(8)  \\
		C0  &  -7(2) &  1.9(4) &  -5.2(3) & 0.34(2) \\
		C0x   &  -23(9) &  5(1) &  -2.7(6) &  0  \\
		C1   &  -11(2) &  2.2(3) &  -3.5(3) &  0  \\
		\enddata
		
		
		\tablecomments{For these fits, a second color was only needed for 3 of the combinations of \textcolor{black}{C/O} composition and pulsation mode to be of sufficient quality.
			If no second color was used, $D=0$.}
		
		
	\end{deluxetable*}
	\label{tab:color_mdot_fits}
	\newpage
	\bibliography{bibliography}{}
	\bibliographystyle{aasjournal}
	
\end{document}